\documentclass[%
reprint,
superscriptaddress,
nofootinbib,
amsmath,amssymb,
aps,
floatfix,
10pt
]{revtex4-2} 
\usepackage[latin1]{inputenc}
\usepackage{amsmath,amssymb}
\usepackage{mathrsfs} 
\usepackage{hyperref}
\usepackage[capitalise]{cleveref}
\usepackage{siunitx}
\usepackage{braket}
\usepackage{calc}
\usepackage{tabularx}
\usepackage{dsfont}
\usepackage{color}
\usepackage{seqsplit}
\usepackage{ifthen}
\usepackage[normalem]{ulem}
\usepackage{graphicx}
\usepackage{mathtools} 

\allowdisplaybreaks

\newcommand{\ii}{\mathrm{i}}%

\newcommand{\dif}{\mathrm{d}}%
\newcommand{\tdif}[2]{\frac{\dif#1}{\dif#2}}%
\newcommand{\pdif}[2]{\frac{\partial#1}{\partial#2}}%
\newcommand{\Nabla}{\vec{\nabla}}%
\newcommand{\fdif}{\operatorname{\delta}}%
\newcommand{\Fdif}[2]{\frac{\fdif\!#1}{\fdif\!#2}}%

\newcommand{\uu}{\hat{u}}

\newcommand{\ZT}[1]{\textquotedblleft#1\textquotedblright}%
\newcommand{\ZTP}[1]{`#1'}%

\newcolumntype{Y}{>{\centering\arraybackslash}X}%
\newcolumntype{Z}{>{\raggedright\arraybackslash}X}%

\newlength{\myl}%
\newcommand{\SUM}[2]{{\setlength{\myl}{\widthof{$\displaystyle\sum_{#1}^{#2}$}*\real{0.5}-\widthof{$\displaystyle\sum$}*\real{0.5}}\sum_{#1}^{#2}\;\hspace{-\the\myl}}}
\newcommand{\INT}[3]{\settowidth{\myl}{$\displaystyle\int_{#1}^{#2}$}{\int_{#1}^{#2}\;\;\;\hspace{-\the\myl}\dif #3}\,}
\newcommand{\TINT}[3]{\settowidth{\myl}{$\int_{#1}^{#2}$}{\int_{#1}^{#2}\!\ifthenelse{\equal{#1#2}{}}{}{\;\;\;\;\hspace{-\the\myl}}\dif #3}\,}%
\newcommand{\EINT}[3]{\settowidth{\myl}{$\int_{#1}^{#2}$}{\int_{#1}^{#2}\;\;\;\,\hspace{-\the\myl}\dif #3}\,}

\usepackage{amsthm}

\newtheorem{definition}{Definition}

\begin{document}
	\title{What exactly is \ZTP{active matter}?}
	
	\author{Michael te Vrugt}
	\email[Corresponding author: ]{tevrugtm@uni-mainz.de}
	\affiliation{Institut f\"ur Physik, Johannes Gutenberg-Universit\"at Mainz, 55128 Mainz, Germany}
	
	\author{Benno Liebchen}
	\affiliation{Institut f\"ur Physik der kondensierten Materie, Technische Universit\"at Darmstadt, 64289 Darmstadt, Germany}
	
	\author{Michael E. Cates}
	\affiliation{DAMTP, Centre for Mathematical Sciences, University of Cambridge, Cambridge CB3 0WA, United Kingdom}

	\begin{abstract}
		As the study of active matter has developed into one of the most rapidly growing subfields of condensed matter physics, more and more kinds of physical systems have been included in this framework. While the word \ZTP{active} is often thought of as referring to self-propelled particles, it is also applied to a large variety of other systems such as non-polar active nematics or certain particles with non-reciprocal interactions. Developing novel forms of active matter, as attempted, e.g., in the framework of quantum active matter, requires a clear idea of what active matter is. Here, we critically discuss how the understanding of active matter has changed over time, what precisely a definition of \ZTP{active matter} can look like, and to what extent it is (still) possible to define active matter in a way that covers all systems that are commonly understood as active matter while distinguishing them from other driven systems. Moreover, we discuss the definition of an \ZTP{active field theory}, where \ZTP{active} is used as an attribute of a theoretical model rather than of a physical system. We show that the usage of the term \ZTP{active} requires agreement on a coarse-grained viewpoint. We discuss the meaning of \ZTP{active} both in general terms and via the specific examples of chemically driven particles, ultrasound-driven particles, active nematics, particles with non-reciprocal interactions, intracellular phase separation, and quantum active matter.
	\end{abstract}
	\maketitle
		
	\section{Introduction}
 
	{\it A common identity--}
	The research field of active matter \cite{MarchettiJRLPRS2013,BechingerdLLRVV2016,Ramaswamy2019,OByrneKTvW2021,ShaebaniWWGR2020,GompperEtAl2020,teVrugtW2024} unifies a broad range of studies in nonequilibrium systems, many of which are centered around self-propelled agents like bacteria, birds or synthetic colloidal microswimmers. It has evolved into one of the major fields of research in physics, and is undoubtedly a topic that many colloquium talks are given on. Speakers giving a general talk about active matter often start their talks with a brief and incomplete explanation of what active matter is -- and may find themselves faced with an unexpectedly difficult question: What is active matter, actually?
	
	Without doubt the term \ZTP{active matter}\footnote{In this work, we use single quotation marks if we are referring to a word rather than to its meaning \cite{Quine1940}. For instance, we would say that Einstein was a physicist and that \ZTP{Einstein} has eight letters -- in the latter case, since \ZTP{Einstein} is in quotation marks, we speak about the word rather than about the person. Double quotation marks are used for citations.} has been useful for the research field: it has served as a common identity for researchers with similar interests and has helped to define what is new about their research activities. The term has also 
	helped researchers to efficiently identify relevant content in scientific journals and to define meeting points at conferences. In fact, the evolution of the term's success is reflected by the number of publications carrying the term \ZTP{active matter} in the title, which has increased from a first few relevant works in 2006-2007 to about 5000 publications in 2023-2024 (according to Google Scholar, 20 April 2025). This growth has occurred on the back of the development of an increasingly broad variety of synthetic experimental setups ranging from autophoretic Janus colloids and droplet swimmers to externally actuated systems like Quincke rollers. While all these are instances of self-propelled particles, there are a broad range of other systems, including a variety of other nonequilibrium materials and biological systems \cite{MannaLSB2022,HallatschekDDDEWW2023,ShankarSBMV2020,BurlaMVAK2019,Liverpool2006}, that are now all commonly classified as active matter. This evolution has repeatedly inspired (or demanded) new definitions of \ZTP{active matter}, as we will discuss in this article. Accordingly, to date, we are in a situation, where no unique and generally accepted definition of \ZTP{active matter} prevails. 
	
	{\it Why should we attempt to define \ZTP{active matter}?--}
	It is quite common in physics to use technical terms where everyone has a shared sense of what they mean, but no one can really pin it down. To some extent, this lies in the nature of physical terminology (an excellent illustration is the discussion of how to define \ZTP{force} in chapter 12 of the Feynman lectures \cite{FeynmanLS1964}). If explicit attempts at a definition are made, however, they can make a significant impact on how the field understands itself (prominent examples are the definition of \ZTP{soft matter} by \citet{deGennes1992} or the definition of a \ZTP{quantum computer} by \citet{DiVincenzo2000}). Active matter physics is a somewhat unusual case here, since a large proportion of the articles on this topic start with some explicit explanation of what the authors take active matter to be. These unsurprisingly often head in a similar direction, but they also differ in important respects. A common shorthand explanation is the claim that active matter consists of self-propelled particles. While self-propelled particles are generally classified as active, there are also many non-self-propelled objects that are considered to fall in this category. Moreover, the idea of self-propulsion forces arose primarily in the study of \textit{dry} models (where momentum is not conserved since the substrate or environment that the particles exchange momentum with is not taken into account). However, the study of \textit{wet} models (that have momentum conservation), where activity can also consist in forces that move a surrounding fluid, also has a long tradition \cite{DombrowskiCCGK2004,PaxtonEtAl2004,JoannyP2009}.
	
	We believe that it is time to critically reflect on existing definitions and to 
	ask what precisely active matter is and how it is different from other nonequilibrium systems that are typically not referred to as \ZT{active}. Is it perhaps already too late to agree on a single definition?
	Clearly, deciding precisely what active matter is, would serve a purpose beyond just the intellectual satisfaction of clarifying terminology. Having a definition of active matter would help to clarify what is new. For instance, non-reciprocal interactions have been studied under different names already for many years \cite{LipowskiFLG2015,SanchezLR2002,Onsager1944,LimaS2006,ClausenHS1998,Hinch1988,CaflischLLS1988,Slack1963}, such that a lack of terminological unification leads to the risk of many phenomena being discovered a second time without anyone noticing. (Note that, as discussed in Section \ref{nonreci} below, not every system with nonreciprocal interactions is active.) Another motivation is certainly that it is helpful for teaching purposes -- explaining to a graduate student or in a colloquium talk what active matter is requires putting the intuitions one may have on this into precise words. Similarly, if one wants to summarize research on active matter in a review article (which is done fairly often \cite{teVrugtW2024}), it is necessary to delineate in some way the research works that deal with active matter from those that do not. A definition could also help to understand if (and which) ongoing activities in different research fields are dealing with identical problems (on some fundamental level) and could hence benefit from an exchange of ideas. An important example is the study of self-propelled medical microrobots \cite{NelsonKA2010,LeeRDS2023,MundacaAFZW2023}, which is a large and rapidly advancing field of research. These microrobots often satisy a standard criterion for being an active particle (they use internal energy to propel themselves), yet the community of researchers studying them is mostly separate from the active matter community and uses a different terminology. Finally, knowing exactly what active matter is, 
	would also be helpful when advancing the research field of active matter into new directions in the future. For instance, there is currently an upcoming interest in quantum active matter \cite{YamagishiHIO2023,TakasanAK2023,ZhengL2023,YamagishiHO2023,NadolnyBB2025} and in this connection it would clearly be useful to understand what distinctive features of an open quantum system would make it active -- and distinct from similar features that are already studied in the literature on open quantum systems. It is difficult to judge whether one has successfully developed a quantum-mechanical active particle if it is not clear what exactly that is supposed to be.
	
	In this article, we review and analyze existing attemps to define \ZTP{active matter}. In doing so, we will survey a number of subfields of active matter physics that have shaped our understanding of this question, such as active nematics, non-reciprocal interactions, and quantum active matter. For graduate students and newcomers to the field, this article provides an introductory answer to the question what the field of active matter physics is about. For specialists, this article explains why this question may be more difficult to answer than is often thought, and thus constitutes an invitation to further discussion.
	
	\section{\label{defining}Defining \ZTP{active matter}}
	\subsection{Why is it not obvious how to define active matter?}
	The perhaps shortest consensus answer to the question \ZT{What is active matter?} is \textit{matter comprising active particles}. But what are active particles? A frequently used understanding is that a particle is active if it has the ability to use energy (from its local environment, or from an on-board resource) to create local motion of itself (\textit{mover}) or its environment (\textit{shaker}). 
	It seems clear that agents like bacteria, birds, and synthetic microswimmers like autophoretic Janus colloids are active in that sense. 
	But are we consistent when we conventionally call shape-asymmetric granulates that show directed motion on uniform vibrated plates \ZT{active} \cite{deseigne2010collective} but not shape-symmetric granulates that show directed motion when being exposed to a vibrating plate with a periodic but asymmetric surface pattern \cite{derenyi1998collective,
		farkas1999transitions}? In fact, the latter setup qualifies as a typical ratchet creating directed transport \cite{reimann2002brownian} and is conventionally attributed to ratchet physics (or the physics of Brownian motors) rather than to active matter physics. In addition, we might wonder, for example, if reaction-diffusion systems that are typically not understood as active systems but that can spontaneously form travelling-wave patterns \cite{Turing1952,CrossH1993,Murray2008} should actually qualify as active matter. This is because the underlying molecules essentially use energy from chemical reactions taking place in their environment (similarly to autophoretic Janus colloids) to create directed motion of the pattern. One might argue that such systems are not active, because the individual molecules that are involved in the system can not propel individually. This would however provoke conflicts with the usage of the word \ZTP{active} for colloidal systems that are made of non-active isotropic building blocks that collectively acquire the ability to self-propel at the many-body-level, e.g. through non-reciprocal interactions \cite{soto2014self,schmidt2019light}. 
	These considerations illustrate -- based on one of the most widely used definitions of active matter --  that it is not obvious which systems actually qualify. Later in this article, we will discuss other definitions of active matter to explore if any of those makes for an easier case. 
	
	In addition, the word \ZTP{active} is used not only to describe a property of physical systems, but also to describe a property of theoretical models. There are many examples for this, such as \ZT{active Fokker-Planck equation} \cite{HerreraS2024}, \ZT{active Langevin equation} \cite{WanJLDLY2025}, or \ZT{active phase-field crystal model} \cite{MenzelL2013}. We will, for specificity, focus here on the term \ZTP{active field theories} \cite{Cates2019}, which refers to partial differential equations (PDEs) providing continuum descriptions of active matter systems. Interpreting the word \ZTP{active} as standing for something like \ZT{converting energy into directed motion} \cite{ZottlS2016,MarconiC2025} -- which is a plausible interpretation in a term like \ZTP{active particle} -- is not possible in the case of \ZTP{active field theory} since PDEs do not generally convert energy into directed motion. An obvious definition would be that, for instan\textit{}ce, an active field theory is a field theory describing active particles. This, however, is not helpful since (a) active field theories are often introduced and studied without a specific microscopic model in mind, which makes a definition that explicitly refers to such a model problematic and (b) phenomena such as the collective motion of humans, which are clearly active, are often described with theories that are typically not called \ZT{active} such as e.g. the diffusion equation in the simplest case. 
	
	It is important to note -- and this is the key message of this review -- that we can sensibly communicate about active matter only when (implicitly) agreeing on taking a coarse-grained perspective. To illustrate this, consider an autophoretic Janus colloid \cite{Stark2018} that self-propels by catalyzing a chemical reaction on part of its surface, which leads to a concentration gradient accross this surface, to which the colloid responds by diffusiophoresis or a similar mechanism. When describing this system at the microscopic (atomistic) level, one can use Hamiltonian equations, and at this level  the molecules contained in the colloidal particles and in the solvent simply counterpropagate in a way that momentum is conserved. That is, directed motion becomes visible only when choosing to focus on the colloidal particle. The following quote form the definition of active matter in Ref.\ \cite{BowickFMR2022} illustrates this further: \ZT{\textit{once the chemomechanical processes that convert fuel into motion are integrated out}, the dynamics of such active entities breaks time-reversal symmetry (TRS) in a local and sustained matter} (our emphasis). Systems can be considered active if there is some useful description in which the energy sources and sinks (the explicit inclusion of which in the description of the system would make the system driven rather than active) are not explicitly considered. For example, a system of granular particles on a vibrated plate can be considered active -- if one just studies the particles -- or as driven -- if one also includes the vibrating plate in the model. Consequently, viewing a system as active cannot be done independently of specifying a certain level of description for the system. This issue will be discussed further in Section \ref{coarsenature} using specific examples from chemically driven active matter.
	
	\subsection{\label{history}Some history}
	It is interesting to trace back the term \ZTP{active matter} to its roots, to see where possible ambiguities of what it refers to might have started. The term \ZTP{active matter} itself is quite old; it has been used, for instance, already by E. Rutherford in 1912 \cite{rutherford1912chemical} in an article entitled \ZT{The chemical effects produced by the radiations from active matter}. Here, though, \ZTP{active matter} stood for \ZTP{radioactive matter}.\footnote{Notably, the potential for a radioactive decay is a stored energy which can impact on motion, so some radioactive systems may qualify as active even under more modern definitions, but no one would think in that way now.} Already closer to the way the term is used nowadays is the notion of an \ZT{active walker} from Ref.\ \cite{freimuth1992active}, where it has been introduced to describe an \ZT{agent that is capable of changing the landscape when walking on it}. Among the first occurrences of \ZTP{active} in the current sense is the (according to Ref.\ \cite{romanczuk2012active}) first definition of an \ZT{active Brownian particle} in Ref.~\cite{schimansky1995structure} from 1995, which states: \ZT{We call Brownian particles with the ability to generate a field active Brownian particles if the produced field self-consistently determines the motion of the particles or defines their rates of chemical reactions.} In the same year, the famous Vicsek model for aligning self-propelled particles was introduced \cite{vicsek1995novel}. This work \cite{vicsek1995novel} refers to \ZT{Self-driven particles [..] including live organisms and the so-called 'molecular motors'} and also refers to cars, motile bacteria and discs floating on air tables \cite{lemaitre1990air}. The mentioning of the latter example is interesting, as it serves as an early example where internally powered agents like cars and bacteria were seen in the same light as externally actuated objects. The first occurrence of the term \ZTP{active matter} within the research field we are aware of was in 2006 \cite{ramaswamy2006mechanics}.
	
	Note, however, that research on physical systems that we would nowadays refer to as active matter has already started before this term was invented. While it is hard to identify which work exactly was the earliest, reasonable candidates are the works by \citet{Przibram1913,Przibram1917} (1913 and 1917) and \citet{Furth1920} (1920), who studied the diffusive motion of microorganisms and found certain deviations from the standard equilibrium theory that one would, in modern language, attribute to active motion. What is nowadays referred to as the Vicsek model was used well before 1995 in the study of fish \cite{Aoki1982,Partridge1982} (1982) and in computer science \cite{Reynolds1987} (1987). Many basic ideas of the modern study of microswimmers were presented by \citet{Purcell1977} already in 1977. And \citet{Schnitzer1993} laid the groundwork for the statistical mechanics of self-propelling particles in 1993 without using the word \ZTP{active} anywhere in the paper. A brief historical discussion of some of these works can be found in Refs. \ \cite{RomanczukBELSG2012,BowickFMR2022}.
	
	If the number of review articles is a good indicator, the largest subfield of active matter physics is the study of medical applications \cite{teVrugtW2024}. The historical origins of this idea go again back quite far, namely to Feynman's suggestion that patients may some day be able to \ZT{swallow the surgeon} \cite{Feynman1961}, which takes the form of a microrobot performing a surgery or delivering drugs, and to the movie \textit{Fantastic Voyage} that shows a miniature submarine traveling through the body \cite{Fleischer1966}. In the context of medical applications, artificial active particles are often referred to as microrobots, microbots, nanorobots, or nanobots. A major aim is to use them for targeted drug delivery \cite{LuoFWG2018,HuGCMQY2020,DasS2024}.
	
	Since the invention of the first synthetic microswimmers in 2004 \cite{paxton2004catalytic}, the number of publications using the term \ZTP{self-propelled-particles} has exploded, and the term has been applied to an increasing number of experiments where individual agents, such as chemically active droplets, showed directed motion in an unbiased environment. Around the same time, the first studies on active nematics \cite{MishraR2006,ChateGM2006,AditiR2002}, for which vibrated granular rods were an important example, were published. Consequently, already in the early days of active matter physics, \ZTP{active} was used to refer to two somewhat different kinds of systems (those with polar order, where the polarity sets a direction of collective motion, and those with nematic order).
	
	During the development of the field, the class of systems that is studied in active matter physics has expanded quite significantly. A very explicit discussion of this phenomenon can be found in Ref.\ \cite{SahaAG2020}:
	\begin{itemize}
		\item \ZT{Throughout the years, particular attention has been given to mechanisms that manifestly break equilibrium physics already at the level of single constituents, as is the case for self-propelled agents such as microswimmers in polar active matter [...]. More recently, subtler manifestations of nonequilibrium activity have taken the spotlight, in particular, those related to the interactions between the active agents, which typically include effective nonconservative forces [...]. A particularly interesting realization of active interactions can occur in mixtures of non-self-propelling scalar active matter. Here, activity manifests itself only through the nature of the effective interactions between different particle species.}
	\end{itemize}
	Thus, while early works on active matter focused on particles that create some motion individually -- often in the form of self-propulsion, but sometimes also in shaking or pumping a surrounding fluid -- more recent studies also include systems where a similar dynamics emerges only at a collective level. The above quote by \citet{SahaAG2020} is from a study of mixtures with \textit{nonreciprocal interactions}. Such systems have been studied already for a long time \cite{Lotka1925,ChowW1987,CaflischLLS1988,CasteraM1989}. However, recent work has taken a novel and very fruitful perspective by studying certain systems with nonreciprocal interactions \textit{as active matter systems}. Similarly, it is increasingly common to classify systems with growing and dividing constituents such as cellular systems as active \cite{HallatschekDDDEWW2023,LishHGB2024,PollackBG2022}. While motile cells are active in the traditional sence, the consideration of \ZT{proliferation [...] as a source of non-equilibrium activity} \cite{LishHGB2024} now seems to qualify also non-motile cells as active matter. 
	
	The story is not over, as new subfields continue to open up in active matter physics. An example for this is the idea of realizing quantum-mechanical active particles. The first works on this topic have been published quite recently \cite{AdachiTK2022,NadolnyBB2025,KhassehWMWH2023,ZhengL2023}. Again, though, difficulties in tracing the origins of this field arise from the fact that earlier (even experimental) realizations, such as quantum nanoparticles with non-reciprocal interactions \cite{RieserEtAl2022,ReisenbauerEtAl2024}, can be thought of as quantum-mechanical active matter, but have not been referred to in this way.
 
	\subsection{\label{existence}Existing definitions}
	An article on the definition of \ZTP{active matter} may seem unnecessary since the literature contains numerous of definitions of this kind -- many articles on active matter start with such a definition, and a superficial look at these definitions seems to suggest that they all have a relatively similar scope. However, in fact the situation is far from clear. To illustrate this, we discuss here a few definitions from major review articles (listed here chronologically):
	\begin{definition} \textbf{(2010)}
		\ZT{The aspect of biological matter of interest here is the ability to transduce free energy into systematic movement. This property is the defining characteristic of active matter (...). The direction of self-propelled motion is set by the orientation of the particle itself, not fixed by an external field. Indeed, these can be taken as the practical defining properties of active matter.} \cite{Ramaswamy2010}
		\label{defa}
	\end{definition}
	\begin{definition}
		\textbf{(2013)} \ZT{A distinctive, indeed, defining feature of active systems compared to more familiar nonequilibrium systems is the fact that the energy input that drives the system out of equilibrium is local, for example, at the level of each particle, rather than at the system's boundaries as in a shear flow. Each active particle consumes and dissipates energy going through a cycle that fuels internal changes, generally leading to motion.} \cite{MarchettiJRLPRS2013}
		\label{defb}
	\end{definition}
	\begin{definition}
		\textbf{(2020)} \ZT{A variety of media can be regarded as active matter, that is, out-of-equilibrium systems composed of individual components that convert energy into non-conservative forces and motion at the microscale} \cite{ShankarSBMV2020}   
		\label{defc}
	\end{definition}
	\begin{definition}
		\textbf{(2021)} \ZT{In active matter systems, the fundamental constituents dissipate energy to exert self-propelling forces on the environment.} \cite{OByrneKTvW2021}    
		\label{defd}
	\end{definition}
	\begin{definition}
		\textbf{(2022)} \ZT{Today, the name active matter refers to any collection of entities that individually use free energy to generate their own motion and forces (...) The defining property of an active system is that the energy input that maintains the system out of equilibrium, whether truly internal or created by contact with a proximate surface, acts individually and independently on each \ZT{active particle.} Hence, once the chemomechanical processes that convert fuel into motion are integrated out, the dynamics of such active entities breaks time-reversal symmetry (TRS) in a local and sustained matter. This should be contrasted with more conventional nonequilibrium systems that are displaced from equilibrium globally by an external force that picks out a direction in space, as in sedimentation under gravity, or are forced at the boundaries, such as through an imposed mechanical shear.} \cite{BowickFMR2022}   
		\label{defe}
	\end{definition}
	\begin{definition}
		\textbf{(2024)} \ZT{Active matter comprises entities that dissipate energy to exert propelling forces on their environment.} \cite{GranekKKRST2024}
		\label{deff}
	\end{definition}	
	\begin{figure}
		\centering
		\includegraphics[width=0.5\linewidth]{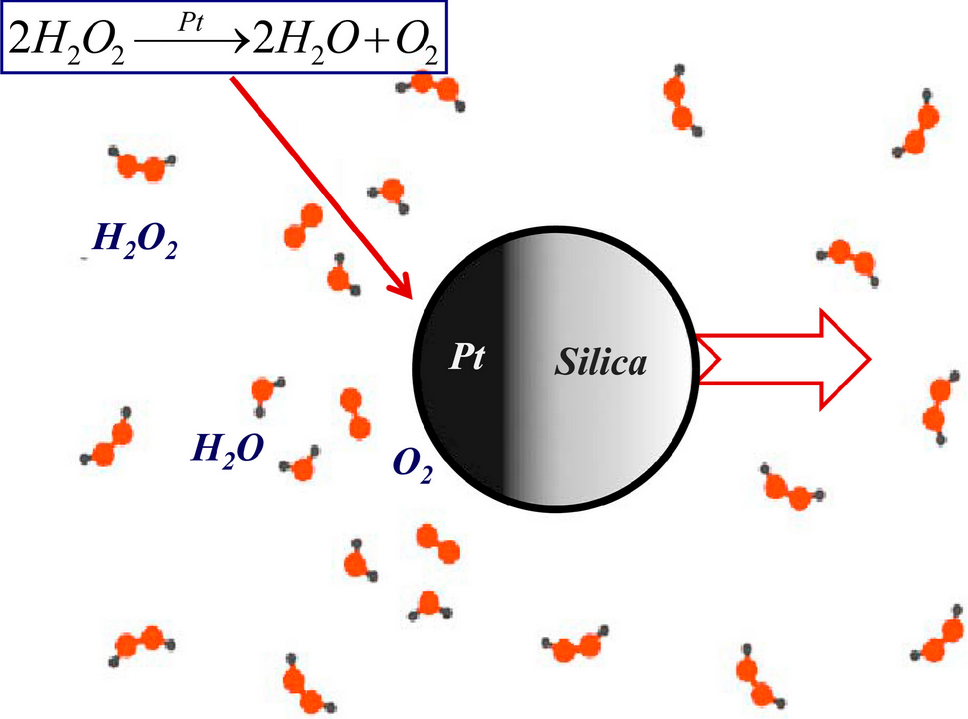}
		\caption{Self-propelling Janus particle made of platinum and silica. The platinum side catalyzes the decomposition of hydrogen peroxide into water and oxygen \cite{ZhangZCS2017}. Reproduced with permission from Ref.\ \cite{ZhangZCS2017}.}
		\label{fig:janus}
	\end{figure}
	These definitions clearly all go in a similar direction and do all classify some paradigmatic examples of active matter as indeed belonging to this class:
	\begin{itemize}
		\item Janus particles \cite{WaltherM2008} -- visualized in \cref{fig:janus} -- have an asymmetric coating, as a consequence of which they catalyze a chemical reaction in a uniform bath on part of their surface only, which leads to a concentration gradient across their own surface. This gradient drives the particle forward (\cref{defd,deff}), in a direction dependent on the particle orientation (\cref{defa}), through diffusiophoresis or a similar mechanism. The force is nonconservative, and the energy influx is local since the chemical reactions take place wherever the particle is (\cref{defb,defc,defe}).
		\item Birds \cite{CavagnaG2014} eat food in order to gain internal energy that they store. This internal energy is then converted into (non-conservative) propulsion forces when the bird flies (\cref{defd,deff}). The direction of motion is determined by the bird's orientation (\cref{defa}). The energy input driving the motion is local, at the level of each bird, and the propulsion force is nonconservative (\cref{defb,defc,defe}).
	\end{itemize}
	One could provide a discussion of this type for many systems studied in active matter physics. However, there are also systems for which these definitions would disagree and whose status as active matter systems is therefore less obvious (and in some cases controversial):
	\begin{itemize}
		\item A Brownian ratchet 
		\cite{reimann2002brownian}, despite not generally regarded an active particle, would probably count as active under \cref{defa}  as it uses free energy, which is transmitted to each particle individually, to create motion. 
		\item Non-self-propelling particles with vision cone interactions \cite{HuangtVWL2024} would satisfy \cref{defb,defc,defe} as they are driven out of equilibrium by a local energy influx leading to non-conservative forces.
		\item Particles that dynamically change their size have often been thought of as active \cite{ZhangF2023,MonchoD2020}, and would certainly count as such by \cref{defb,defe}.
		\item Mixtures of two particle species driven by two different thermostats are sometimes viewed as active and show segregation behavior reminiscent of that of active particles \cite{GrosbergJ2018,GrosbergJ2015}. They are driven out of equilibrium by a local energy influx, as demanded by \cref{defb,defe}. (It is assumed here that the thermostats couple to each particle individually, particles in contact with a hot and a cold wall would not have a local energy influx.)
		\item Growing and dividing systems are sometimes classified as \ZTP{proliferating active matter} \cite{HallatschekDDDEWW2023,LishHGB2024,PollackBG2022,MitchellT2022}. Such systems satisfy \cref{defb} since they are driven out of equilibrium by a local energy influx.
	\end{itemize}
	It is, however, unclear whether any of these systems satisfied \cref{defd,deff}, i.e., whether they generate propulsion forces.
	
	\subsection{\label{howtofindit}How to find definitions}
	This short survey of existing definitions of \ZTP{active matter} has shown that there are notable differences in how different authors understand this concept, and that physicists thus can (and do!) have different views regarding the question whether a certain system is active. This is not intrinsically problematic, as definitions are to a large extend conventions that are not right or wrong. It is not the aim of this article to criticize or correct any of the definitions shown so far -- they are all reasonable reflections of a broad class of research works in the field. 
	
	In general though, one can offer a definition of a term like \ZTP{active matter} with two different kinds of intentions: First, it can be \textbf{descriptive} in the sense that we describe what people \textit{do} mean when they say \ZTP{active matter}. Second, it can be \textbf{prescriptive}, meanig that we say what people \textit{should} mean when they say \ZTP{active matter}. In this work, we are concerned with formulating improved descriptive definitions by extracting what researchers actually do mean by \ZTP{active matter}. The difficulty here is that, as the above summary has shown, different researchers may mean different things. Moreover, if a word is used in a way that had some historical development, it can be hard to then later put this historical development into a fixed definition that is not reflective of it. Thus, while many researchers have an intuitive idea of what an active particle is, it can be difficult to precisely fit this into words.\footnote{An excellent example is Plato`s definition of a human: According to an urban legend \cite{Laertius2018}, Plato received much applause for defining a human as a \ZT{featherless biped}, until Diogenes plucked a chicken and brought it, with the words \ZT{Here is Plato's man}, into the academy. (The definition of \ZTP{human} was subsequently changed to \ZT{featherless biped with broad flat nails}.)} A general issue here is the question what to do if one has settled on a definition of, say, \ZTP{active matter}, and then finds that it includes a system that one would not intuitively think of as being active (or that it does not include a system that one wants to call active). In this case, one can either decide to accept the definition and to from now on (or no longer) also classify this system as active, or one can use this result as an indication that the definition should be changed. 
	
	If physicists decide to extend the use of \ZTP{active} towards new systems but also wish to avoid it becoming so broad that it ceases to single out a very special kind of materials, they may at some point be forced to give up the idea that the term \ZTP{active matter} can be given a clear definition. In philosophy of language, the term \textit{family resemblance} \cite{Wittgenstein1953} is commonly used to characterize this phenomenon, with a typical example being the word \ZTP{game}: There are a variety of things that this word is used for (board games, card games, ball games, ...) that all have something in common, but it is not really possible to define this word in the sense that one specifies necessary and sufficient conditions that are satisfied by everything that is a game and by nothing else. The term \ZTP{active matter} is nowadays used in a similar way -- every new system that is called \ZTP{active matter} has something in common with other things that were already called \ZTP{active matter}.
	
	\section{\label{modern}Active matter as a special class of nonequilibrium systems}
	
	\subsection{\label{trsandep}Time-reversal symmetry and entropy production}
	
	To see why one may want to refer to, e.g., nonreciprocally interacting particles as \ZTP{active} and to thereby make them part of a classification that initially was thought to contain primarily self-propelled objects, it is useful to note that self-propelled particles have a very interesting property: Their dynamics violates \textit{time-reversal symmetry} (TRS). If one records a video of a system of Brownian particles in equilibrium and plays it backwards, the result would (according to the Langevin equations governing their dynamics) be an equally likely process as the one that has actually occurred. Therefore, the dynamics of this system is said to obey TRS \cite{OByrneKTvW2021}, i.e., when considering an unbiased ensemble of initial conditions, forward and backwards processes have the same statistical properties (averages). If such a system is in a stationary state, there is no way of distinguishing whether such a process is played forwards or backwards. For active particles, the situation is different. Take as an example a group of humans walking through the street. If you record a video of this process and play it backwards, you will see a group of humans walking backwards, which is a relatively unlikely process.
	
	This observation is quantified by the systems \textit{informatic entropy production}. The idea is that one can measure the degree of TRS breaking in a certain process by assessing how likely this process is compared to its time reversal. In the case of an equilibrium system, these probabilities are equal. Therefore, the \textit{steady-state path entropy production}\footnote{We provide here the definition for a continuous time-process. There are several ways of generalizing \cref{entropyproduction} to the discrete-time case, see Ref.\ \cite{LeeLP2024} for a discussion.} \cite{Seifert2012}
	\begin{equation}
		\mathcal{S} = \lim\limits_{t\to\infty}\frac{1}{t}\left\langle\ln\frac{P_\mathrm{F}(\Psi(t))}{P_\mathrm{B}(\Psi(t))} \right\rangle
		\label{entropyproduction}
	\end{equation}
	with time $t$, forward probability $P_\mathrm{F}$, backward probability $P_\mathrm{R}$, and generic state variable $\Psi$, is zero. That is, if the system has reached a steady state (thermodynamic equilibrium), forward and backward probabilities are equal. In contrast, for systems of self-propelled particles, the entropy production in the steady state is in general positive. The example of humans walking forwards and backwards provides an intuitive explanation for why, as one can easily see why in this case the backwards case has a low probability. A steady state is a state where the system has a stationary probability measure (it may still exhibit currents or time-dependent phenomena in this state).
	
	The informatic entropy production \eqref{entropyproduction} is to be contrasted with the \textit{thermodynamic entropy production}, which is a measure for the degree of energy dissipation in the system. In general these two entropy productions are different -- the full thermodynamic entropy production of a group of humans is unrelated to the probability of forward or backward walking. However if applied to a  microscopic description containing all degrees of freedom relevant for energy dissipation, \eqref{entropyproduction} does represent the thermodynamic entropy production as studied in stochastic thermodynamics \cite{Bebon2024}.
	
	An important factor to take into account in the context of informatic entropy production is the fact that it depends on what one takes to be a time-reversal symmetry. This is particularly relevant in the context of activity, which arises only when taking a coarse-grained point of view. (A more technical discussion of this point can be found in Ref.\ \cite{FodorLC2021}. The definition of TRS is discussed in Ref.\ \cite{DieballG2025}.) Take the example of humans given above, where we have argued that their behavior violates TRS (walking backwards is unlikely). We could also have insisted that, when playing the movie backwards, we have to rotate the bodies of the humans by an angle $\pi$ -- in other words, we could have changed the time reversal symmetry operation. After all, we cannot really change the direction in which time flows, such that what we take to be a time reversal depends on more indirect arguments like a certain equilibrium limit \cite{FodorLC2021} and is thereby to some extent a subjective choice. A similar issue arises in electromagnetism \cite{Malament2004,Albert2000}: Whether the equations of motion of a particle in a magnetic field are time-reversal invariant depends on whether or not one assumes that the magnetic field changes its sign under time reversal, and the reason why one usually assumes that it does is that this assumption ensures that Maxwell's equations are invariant under time reversal. Note that even for a passive system with a well-defined temperature, the entropy (production) from classical thermodynamics coincides with that from \cref{entropyproduction} only when the correct time reversal symmetry operation is chosen.
	
	Moreover, entropy production also depends on the degrees of freedom one takes into consideration. For instance, we could have recorded a really blurry video in which we cannot really see in which direction the humans look. In this case, we could not have seen whether they walk forwards or backwards. And in this case, we might have been unable to distinguish between the forward and the backward video, inferring that the system exhibits TRS. In other words: The (informatic) entropy production depends on the degrees of freedom we consider \cite{BilottoCV2021,FodorLC2021,ShankarM2018}.
	
	It is helpful to explicitly illustrate these ideas using a simple particle-based dry model for an active particle. Here, the position $\vec{r}_i$ in a many-particle system are governed by the stochastic differential equations
	\begin{equation}
		\dot{\vec{r}}_i  = \beta D(\vec{f}_i - \Nabla_{\vec{r}_i}U(\{\vec{r}_j\})) + \sqrt{2D}\vec{\xi}_i	\label{langevin2}
	\end{equation}
	with translational diffusion coefficient $D$, thermodynamic $\beta=1/(k_\mathrm{B}T)$ with Boltzmann constant $k_\mathrm{B}$ and temperature $T$, potential $U$, self-propulsion force $\vec{f}_i$, and Gaussian white noises $\vec{\vec{\xi}_i}$ that have zero mean and unit variance. Important special cases of \cref{langevin2} are the active Brownian particle \cite{RomanczukBELSG2012} and the active Ornstein-Uhlenbeck particle \cite{MaggiMGDL2015,MartinEtAl2021}. When computing the informatic entropy production of active particles, one can find three different results depending on which degrees of freedom one tracks and how the TRS operation is defined \cite{FodorLC2021}:
	\begin{enumerate}
		\item We account only for the particle positions. In this case, \cref{entropyproduction} gives
		\begin{equation}
			\mathcal{S}= - 4\beta D_\mathrm{T}\INT{-\infty}{\infty}{t}\Gamma(t)\sum_{i}\braket{\dot{\vec{r}}_i\cdot\Nabla_{\vec{r}_i}U},
			\label{epr1}
		\end{equation}
		where $\braket{\cdot}$ denotes an ensemble average and where the function $\Gamma(t)$ depends on the statistics of $\vec{f}_i$. Note that \cref{epr1} gives zero for $U=0$. In the example of the humans, \cref{epr1} is the scenario where we do not resolve in which directions they look.
		\item We account for positions and orientations and assume that the orientation flips sign under time reversal. In this case, \cref{entropyproduction} gives
		\begin{equation}
			\mathcal{S}=\frac{\beta D_\mathrm{T}}{T}\sum_{i}\braket{\vec{f}_i\cdot\Nabla_{\vec{r}_i}U}.
			\label{epr2}
		\end{equation}
		In the human video example, this would be a scenario where the humans look in the opposite direction when we play the video backwards, which would for $U=0$ not allow us to find a difference from the forward case -- consequently, \cref{epr2} vanishes for $U=0$.
		\item We account for positions and orientations and assume that the orientation does not flip sign under time reversal
		\begin{equation}
			\mathcal{S}=\frac{1}{T}\sum_{i}\braket{\vec{f}_i\cdot\dot{\vec{r}}_i}.
			\label{epr3}
		\end{equation}
		This formalizes the idea of a video of people walking backwards discussed at the beginning of this section. In this case, the informatic entropy production is nonzero for $U=0$, since the time-reversed dynamics (humans walk backwards) is unlikely.
	\end{enumerate}
	
	An interesting observation that can be made concerning the relation of TRS and irreversibility is the following: There is a long-standing debate in the foundations of physics \cite{Zeh1989,Albert2000,teVrugt2020,DingC2025} concerned with the question why the macroscopic world has a clear arrow of time -- manifest in the irreversible approach to thermodynamic equilibrium -- given the fact that the microscopic dynamics, which has a Hamiltonian form, is time-reversal invariant. Explicit coarse-graining of this reversible dynamics does give irreversible macroscopic equations that describe an approach to equilibrium \cite{teVrugt2022}. Active matter physics adds an interesting twist here since it shows that systems that do \textit{not} have TRS on the microscopic level do in fact not approach equilibrium. In other words, rather than being in conflict with the approach to equilibrium, TRS seems to be a prerequisite for it. More technically, this can be seen from the fact that microscopic TRS implies reciprocity relations for macroscopic transport coefficients, which are then used in showing that the system approaches equilibrium on a macroscopic level \cite{Onsager1931}.
	
	\subsection{Classifying nonequilibrium systems}
	Self-propelled particles like humans are among many system that can exhibit a non-zero informatic entropy production in the steady state. Two further examples would be
	\begin{itemize}
		\item a water slide where water is flowing from the top to the bottom and then pumped back up -- the time reversal would be water freely flowing from bottom to top and then being pumped downwards, which is very unlikely.
		\item a colony of (non-moving) bacteria that grow, divide and die, in a way that the colony has reached a steady state \cite{CameronT2025} -- the time reversal would be bacteria that emerge spontaneously, shrink, and merge, which is very unlikely.
	\end{itemize}
	The group of humans, the water slide, and the bacteria have in common that an energy influx keeps them out of equilibrium. However, there is an important respect in which the humans and the bacteria differ from the water slide: In the latter case, the individual water molecules are perfectly ordinary passive particles, their motion is solely the result of an \textit{external} force. The humans and the bacteria, in contrast, exhibit TRS breaking due to an \textit{internal} energy supply.
	
	Such differences can be used as a basis for classifying nonequilibrium systems into different classes that can substantially differ in their nonequilibrium behavior:
	\begin{enumerate}
		\item Nonequilibrium systems that are approaching equilibrium. An example is an isolated system consisting of two parts at different temperatures -- after a sufficiently long time, their temperatures will be equal.
		\item Nonequilibrium systems that do not approach equilibrium because they are kinetically arrested. An example would be a supercooled liquid below the glass transition which is so dense that the particles cannot reach their equilibrium crystal positions due to caging -- no equilibrium, but not because there is an external or internal drive.
		\item Nonequilibrium systems that are driven by energy influx from the boundaries. A prominent example is a sheared system.
		\item Nonequilibrium systems that are driven by energy influx at the level of each particle. An example would be a swimming bacterium that uses internal energy from nutrients it has consumed.
	\end{enumerate}
	A typical motivation for introducing active matter as a separate class of materials is to distinguish between the third and the fourth variant. There are externally driven and internally driven systems, and the latter class is active. This is especially obvious in \cref{defb}, which notes energy influx at the particle level as a \ZT{defining feature} of activity. However, the distinction between driven and active systems is not always very sharp \cite{Menzel2015}. Moreover, it is hard to argue why a particle in an external field should not have a local energy influx, yet one would not in general think of them as active. These issues will be discussed further in Section \ref{whatitis}.
	
	While distinctions between globally and locally driven nonequilibrium systems and the idea of \ZTP{active matter} as a name for the latter class comes from soft matter research, there is no reason to assume that such distinctions cannot be fruitfully applied also in other areas of physics. An idea in this direction is the emerging research area of quantum active matter (cf. Section \ref{aqm}). In fact, the word \ZTP{active} is already used in fields of physics other than soft matter in a way that is very much in line with this definition -- prominent examples would be \textit{active electrical elements} such as a transistor. Transistors are certainly not self-propelled particles, but they are constituents (particles) of an electrical circuit whose characteristic behavior depends on the existence of a local energy influx (in this case an electrical one).
	
	\subsection{\label{classifythem}Classifying forces}
	Many definitions of active matter rely on mechanical concepts -- in particular concerning forces (see, e.g., the appearance of \ZTP{non-conservative force} in \cref{defc}). Therefore, it is helpful to clarify at this point a variety of concepts arising in the context of discussing forces in active matter:
	\begin{itemize}
		\item \textbf{Wet and dry models} \cite{MarchettiJRLPRS2013}: In a dry model, the medium surrounding an active particle (in many cases a fluid) is not modeled explicitly, such that the overall momentum in the model is not usually conserved and the forces (excluding any explicit external forces) do not have to add up to zero. In a wet model, on the other hand, one describes the dynamics of both the active particles and the surrounding fluid, implying that momentum has to be conserved in the description. Strictly speaking, \ZTP{wet} and \ZTP{dry} refer to different types of models rather than to different types of systems since violations of momentum conservation always result from momentum exchange with an environment that one could in principle also model. However, there are systems (such as granular rods on a surface) where a dry description is very natural, and also systems (such as microorganisms exhibiting bioconvection \cite{PedleyK1992}) where the fluid dynamics is very important and for which a wet model is therefore appropriate. In this sense, one can in practice use \ZTP{wet} and \ZTP{dry} also as attributes of physical systems.
		\item \textbf{Self-propulsion force:} This force appears in the equations of motion for a self-propelled particle in a dry model (the $\vec{f}_i$ in \cref{langevin2}). Notably, it is in fluid mechanics somewhat controversial whether this should actually be referred to as \ZT{force}. On the one hand, one would measure a force if one attaches a force gauge to a self-propelled colloidal particle. On the other hand, the flow field around a colloidal particle that is dragged through a fluid by an external force is different from the flow field around a colloidal particle that is self-propelled, since the latter is actually force-free \cite{ZoettlS2022}. A more detailed discussion of this issue can be found in Ref.\ \cite{tenHagenWTKBL2015}. In a nutshell, the notion of a self-propulsion force arises on an effective level in a dry description that arises from coarse-graining a more microscopic wet description.
		\item \textbf{Nonconservative force:} In general, a conservative force field is one that can be expressed as the gradient of a potential, has zero curl and does no work along a closed path (these criteria are equivalent on a simply connected domain) \cite{WittkowskitV2024}. These critera are usually violated for, e.g., self-propulsion forces, suggesting nonconservative forces as a necessary criterion for activity. However, non-conservativeness is surely not sufficient -- for instance the Lorentz force on a charged particle is nonconservative (although when field energies are included, energy is conserved). If one thus (as done, e.g., in \cref{defc}) maintains that activity manifests itself in nonconservative forces, one needs to be aware that these forces represent a subset of all nonconservative forces, namely those arising from a local influx and conversion of energy.
		\item \textbf{Nonreciprocal interaction force:} These are forces violating Newton's third law, i.e., the relation $\vec{F}_{ij}=-\vec{F}_{ji}$ with $\vec{F}_{ij}$ being the force from particle $i$ on particle $j$ is not valid. This implies a violation of momentum conservation. At least formally, it is not necessary that nonreciprocal interaction forces are nonconservative since momentum and energy conservation are independent requirements resulting from different symmetries. Consider, e.g., the force derived from the interaction potential 
		\begin{equation}
			U_2 (\vec{r}_1,\vec{r}_2)= -a\vec{r}_1^2e^{-\frac{\vec{r}_2^2}{b}}
		\end{equation}
		with constants $a$ and $b$ and positions $\vec{r}_1$, $\vec{r}_2$ of particles 1 and 2. This potential violates translational invariance, and we have nonreciprocal but conservative interactions ($\vec{F}_{1}+\vec{F}_2\neq \vec{0}$ with $\vec{F}_i = \partial U/\partial \vec{r}_i$).
		\item \textbf{Transverse interaction forces:} \cite{LangerSMK2024,GhimentiBSvW2023,GhimentiBSvW2024} These are interaction forces between two particles where the force vector is perpendicular to the vector separating the particles. In the case of overdamped Brownian motion, this is equivalent to having an odd mobility where the velocity of a particle acquires a component perpendicular to the applied force. Such forces or mobilities are not by themselves an indication that the system of interest is active or out of equilibrium. For instance, the Thiele equation \cite{Thiele1973} governing the dynamics of magnetic skyrmions \cite{GeEtAl2023,BremsEtAl2025} generally gives rise to this phenomenon. However, there are also examples of active transverse interactions that lead to spontaneous rotation of initially resting particles around each other \cite{TanMLCHFGDF2022,HuangtVWL2025}. This corresponds to an effective nonreciprocal interaction, specifically to a nonreciprocal interaction torque.
	\end{itemize} 
	
	The last aspect links to an interesting general point, namely the relation between active and magnetic systems. While a notable aspect of active systems is the fact that the breaking of time-reversal invariance of their dynamics allows for persistent currents in steady state, a steady state with persistent currents is also possible in a magnetic system. Therefore, there are many shared phenomena that link the physics of active matter with that of magnetic systems, such as odd dynamics \cite{BanerjeeSAV2017,KalzVASLS2022} or Hall effects \cite{LouEtAl2022,SiebersBJS2024}. Still, the dynamics of a passive particle in a magnetic field \textit{is} invariant under time reversal (provided that one assumes that the magnetic field changes sign under time reversal, as is usually done \cite{Malament2004}), such that magnetic systems are -- unlike active systems -- not necessarily out of equilibrium. This has to do with the fact that the currents in an active system in steady state are dissipative, whereas in a magnetic system this is not (necessarily) the case. Dissipative currents give (thermodynamic) entropy production in and of themselves, so if they are included in a coarse grained description and present in steady state, the system must be active or at least driven. Consequently, comparing active and magnetic systems provides an interesting way of finding out whether a particular effect in active matter necessarily requires a nonequilibrium system, or whether it may also occur in an equilibrium setup. 
	
	Good examples here are the phenomena of \textit{odd viscosity} and \textit{odd elasticity}, which correspond to violations of certain symmetries that the viscosity and elasticity tensor, respectively, usually have (see Ref.\ \cite{FruchartSV2023} for a review, for odd elasticity see also Section \ref{nonreci}). Both phenomena have been found in a variety of active matter systems \cite{BanerjeeSAV2017,MarkovichL2021,ScheibnerSBSIV2020,FossatiSFV2024}. However, odd viscosity is a consequence of microscopic TRS breaking whereas odd elasticity results from microscopic nonconservative forces \cite{FruchartSV2023}. Consequently, odd viscosity has also been found in many non-active and fully Hamiltonian systems \cite{AvronRZ1995,GaneshanA2017,Avron1998,FruchartSV2023} and is not necessarily a non-equilibrium phenomenon. Odd elasticity, in contrast, allows one to set up a cyclic process consisting of quasistatic deformations where nonzero work is done, implying that there has to be an energy source \cite{ScheibnerSBSIV2020}.
	
	\section{\label{whatitis}What active matter is}
	As discussed in Section \ref{history} above, over the years more and more systems have been classified as active matter. This can, from a terminological perspective, in principle have two reasons: Either the definition has become broader, or researchers have come to realize that more systems fall under a definition that was previously designed for a relatively narrow class. In practice it is probably a combination of both factors, but if one looks at the definitions presented in Section \ref{existence}, then the understanding stays relatively consistent. The main difference is that some definitions (\cref{defd,deff}) require self-propulsion while others do not, but this is not a historical development -- also older definitions such as \cref{defb} do not explicitly restrict themselves to self-propelled particles, and indeed the study of active nematics (which are not in general self-propelled) started around a similar time as the study of self-propelling active matter. Moreover, the oldest definition (\cref{defa}) explicitly considers biological matter. This, however, should not be taken as stating that all active systems are necessarily biological, it simply manifests the fact that the idea of active matter originated from the study of biological systems and that the development of synthetic materials exhibiting activity started more recently. We thus here extract, from the definitions presented in Section \ref{existence}, the main features that a system is required to have to count as active.
	
\begin{figure*}
\hspace*{-1.5cm}
\begin{tabular}{ |c|c|c|c|c|c| } 
	\hline
	\textbf{System} & \textbf{Out of equilibrium} & \textbf{Local energy influx} & \textbf{Motion and forces } & \textbf{Directionality}\footnote{We consider this criterion to be satisfied if there is an orientational asymmetry that would not exist without the particle. It is not satisfied if the considered system either involves an external field that breaks the symmetry even in the absence of the relevant particle or if the particles induce only isotropic forces (as for a single pulsating particle).} & \textbf{Viewed as active}\\ 
	 &  & & \textbf{generated by influx} & &\textbf{in the literature}\\ \hline
	Birds \cite{CavagnaG2014}& Yes & Yes & Yes & Yes & Yes\\ \hline
	Janus particles \cite{WaltherM2008} & Yes & Yes & Yes & Yes & Yes\\ \hline
	Vibrated granular particles \cite{NarayanRM2007} & Yes & Yes & Yes & Yes & Yes\\ \hline
	Wet active nematics \cite{DoostmohammadiIYS2018} & Yes & Yes & Yes & Yes & Yes\\ \hline
Self-propelled droplets \cite{Michelin2022} & Yes & Yes & Yes & Yes & Yes\\ \hline
	Ultrasound-driven particles \cite{WangCHM2012} & Yes & Yes & Yes & Yes & Yes\\ \hline
	Charged particle  & Yes & Yes & Yes & No & No\\ 
	in time-dependent electric field &  &  &  & & \\ \hline
	Sedimenting colloids  & Yes & Yes & Yes & No & No\\ \hline
Spring network  & Yes & Yes & Yes & No & Yes\\ 
with odd elasticity \cite{ScheibnerSBSIV2020}  &  &  & &  & \\ \hline
Spherical colloids with  & Yes & Yes & Yes & On many-body level & Sometimes\\ 
multi-species nonreciprocity  &  &  &  & & \\ \hline
Brownian Particles with & Yes & Yes & Yes & Yes & Sometimes\\
vision cone interactions \cite{BarberisP2016,HuangtVWL2024} &  &  &  & & \\ \hline
Nonreciprocal & &  &  &  & \\ 
lattice models \cite{LoosKM2023,AvniFMSV2023} & Yes & Yes & No & Unclear & Sometimes\\ \hline
Brownian ratchets \cite{reimann2002brownian} & Yes & Yes & Yes & No\footnote{Directionality is not satisfied here, because the applied forces locally pick out a direction of space at each time-instance, even after averaging over many initial conditions.} & No\\ \hline
Particles coupled to  & Yes & Yes & Yes & No & Yes\\ 
different heat baths \cite{GrosbergJ2015,SmrekK2017} &  &  &  &  & \\ \hline
	Pulsating particles \cite{ZhangF2023} & Yes & Yes & Yes & On many-body level & Yes\\ \hline
Growing and dividing particles \cite{HallatschekDDDEWW2023} & Yes & Yes & Yes & Sometimes & Yes\\ \hline
Hot Brownian motion \cite{RingsSSCK2010} & Yes & Yes & Yes & No & Sometimes\\ \hline
 Brownian particles & Yes & Yes & Yes & Yes & No\\ \
with time delay \cite{KoppK2023,TaramaEL2019} &  &  &  & & \\ \hline
Chemically driven  & Yes & Yes & Yes & No & Yes\\
liquid-liquid phase separation \cite{JulicherW2024} &  &  &  &  & \\ \hline
	Spins with  & Yes & Yes & Yes & Yes & Yes\\
	directional hopping \cite{SolonT2013,AdachiTK2022} &  & &  &  &\\\hline
	Quantum wavefunction & Yes & Yes & Yes & Yes & Yes\\ 
	dragged along &  & &  &  &\\ 
	AOUP trajectory \cite{ZhengL2023} &  & &  &  &\\ \hline
\end{tabular}
\label{criteria}
\caption{Commonly used criteria for activity illustrated using several examples.}
\end{figure*}

A first key requirement that has been extensively discussed in Section \ref{modern} above is that active matter is \textbf{out of equilibrium}. Many of the spectacular collective phenomena that active matter systems exhibit are consequences of the fact that they are not constrained by the requirements of equilibrium thermodynamics. Notably, active systems should -- as discussed in Section \ref{modern} above -- be permanently (not just transiently) out of equilibrium, also when they are in a steady state. In practice, active systems will have this property only on certain observational timescales since they often require some kind of fuel, and thus they are active only until the fuel runs out. Moreover, if being out of equilibrium is part of the definition of an active particle, there remains the question how we judge whether a certain system is in fact out of equilibrium. This is not always easy to do in an objective way, as will be discussed in Section \ref{coarsenature} below.

Second, there is \textbf{local energy influx}. This is typically why the system is out of equilibrium so that the first and second criteria are not independent. The continous influx and conversion of energy implies that the configuration space of the system needs to host net fluxes of some sort, and equilibrium systems do not exhibit such fluxes. As discussed in Section \ref{modern}, the motivation for this criterion is to distinguish active matter from other nonequilibrium systems that are driven at the boundaries (\cref{defb} makes this very explicit). A local influx can come both from within the particle (as in a bacterium or bird who moves by energy gained from food) or from the particles' local environment (such as in a granular rod driven by a shaking substrate). This distinction is, however, not in general easy to make \cite{Menzel2015}. For instance, if a fluid is heated from below and therefore exhibits Rayleigh-B{\'e}nard convection \cite{Benard1900,Rayleigh1916}, one would certainly consider it driven rather than active since the energy influx comes from the boundary. However, if one heats up a fluid layer locally with a suitably applied laser -- which leads to the same pattern -- one would have a local energy influx. Still, the resulting system would probably not be considered active since it is essentially still exhibiting standard Rayleigh-B{\'e}nard convection.  
	
Third, there are \textbf{motion and forces generated from the local energy influx}. In the simplest case every particle has, in its equation of motion (for example \cref{langevin2}), a self-propulsion force that makes it move forward even in the absence of external forces. As discussed in Section \ref{modern} above, this idea is problematic in the case of particles swimming in a fluid which are actually force-free. Still, active particles generally create some motion of themselves and/or their environment. This can be directed propulsion, but it can also be a systematic change in size (growth) or an oscillation which causes forces on surrounding particles or a surrounding fluid which is transported as a consequence of this. Of importance in this context is the distinction between \textit{dry} and \textit{wet} models discussed in Section \ref{modern} above. In a dry model, momentum is not conserved since momentum sources and sinks are not incorporated. This may create the impression that a particle can propel, in a Munchhausen style\footnote{Baron Munchhausen is a fictional character \cite{Raspe1785} (loosely based on a historical figure) who is known for his tall tales, one of which involves that he gets himself out of a swamp by pulling on his own hair \cite{Burger1786} in violation of momentum conservation.}, by a force it creates itself. Of course this is not actually what happens -- as discussed in Section \ref{classifythem}, these self-propulsion forces arise in a coarse-grained description where flow fields are not explicitly considered. On this level, it looks as if an external force $\vec{f}_i$ is applied to each particle in a direction set by the particles orientation. Definitions explicitly requiring self-propulsion forces (\cref{defd,deff}) are thus clearly working with a dry picture. While external forces and (effective) self-propulsion forces produce the same terms in a dry model equation such as \cref{langevin2}, they give rise to different flow fields \cite{KuemmeltHWTBEVLB2014}.

In a wet model, on the other hand, momentum is conserved and the primary effect of activity is often that an active stress created by the particles moves the surrounding fluid. It is sometimes assumed (e.g., in \cref{defc}) that the forces that an active particle creates should be nonconservative. This is, however, not immediately obvious. For instance, one could build a shaker with an internal battery that creates, say, a time-periodic reciprocal interaction force on its environment. This shaker would presumably count as an active particle, however, there is nothing nonconservative about the force it creates unless one considers also dissipation. What matters is whether the force does work (on the fluid) over the course of a cycle. If so, it is active.
	
Importantly, all definitions require that the local energy influx has to be used to generate the motion (and the forces), i.e., it is formally not sufficient to have a particle that has a local energy influx and that also moves. This implies that it is not sufficient for activity if the local energy influx powers some local chemical reactions or conformational changes (such as nonequilibrium spin flipping in a nonreciprocal lattice model \cite{AvniFMSV2023}) unless these have mechanical effects. Some definitions (such as \cref{defc}) state that an active particle should convert energy into forces, which is a bit imprecise (energy is converted into other forms of energy, not into forces). Moreover, some definitions (\cref{defb,defd,deff}) require that energy is \textit{dissipated}, whereas others (\cref{defc,defe}) simply require that it is converted. This would affect the classification of a particle that converts energy to create directed motion without dissipative losses  would still be active. (Of course if it does so in a fluid the mechanical energy would thereafter be dissipated via viscous heating or similar effects.) As an extreme example, consider a microswimmer that moves through an ideal superfluid. There is no dissipation in this system, and yet most physicists would consider it active.
	
As noted already in Section \ref{modern}, the local energy influx requirement is also satisfied, e.g., by a particle that receives energy from an external field. This energy influx also leads to motion and forces. A criterion that nevertheless sets these particles apart from genuinely active ones is \textbf{directionality}, by which we mean that the direction of the motion or forces that the active particle creates is set by an orientational asymmetry that would not exist without the particle. This is what \cref{defa} refers to as the \ZT{practical defining property of active matter} \cite{Ramaswamy2010} and what also \cref{defe} uses to distinguish active particles from, e.g., sedimenting particles. The directionality requirement also excludes Brownian ratchets, where the fact that a directed motion exists hinges on breaking of parity symmetry by the ratchet potential (although spontaneous ratchets can also exist \cite{reimann2002brownian}). Note that the somewhat complicated formulation \textit{asymmetry that would not exist without the particle} aims to capture the possibility that the particle is not itself asymmetric as a Janus particle is, but simply creates an asymmmetry in its local environment that then leads to particle motion. Importantly, the directionality criterion is also applicable to non-polar active nematics where the particles do not self-propel (see also Section \ref{nemat}). Here, the average direction of the symmetry axis (i.e., the local director) of the particles usually determines the direction in which, e.g., fluid in a wet active nematic is transported. The wet case also shows that directionality does not need to manifest itself in vectorial form. Instead the directionality of a swimming object is defined via a force multipole which is a tensorial object. In practice, the directionality requirement is not always considered essential, as shown both by studies of \ZT{symmetric active colloids} \cite{Golestanian2009} -- for which particles that dynamically change their size \cite{ZhangF2023,MonchoD2020} are a good example -- and by the fact that the definitions using it (\cref{defa,defe}) tend present it more as a rule of thumb for identifying active systems than as a strict requirement in their wording.
	
	Note that the directionality requirement could force one to classify particles as active that one may not otherwise want to classify that way: In the context of Brownian ratchets (not usually considered active), there are examples \cite{MukhopadhyayLS2018} where the drift direction of the colloidal particles that are exposed to the ratchet potential depends on properties of the particle, like size, shape or mobility. When accounting for hydrodynamic interactions between colloidal particles that are exposed to the ratchet, they would show nonreciprocal interactions (similarly to sedimenting spheres), but the (mean) direction of motion would not be determined by the applied external fields alone -- it would also depend on the particle properties. If this particle property is the particle's orientation, it would probably count as active by this criterion.
	
	Having discussed the typical requirements for activity, it is worth noting that there is a number of edge cases, which are considered active in parts of the literature without (obviously) satisfying all of those requirements. A prime example is that of nonreciprocal lattice models, where the particles do not move or create transport forces on their environment, but which do maintain a nonequilibrium state by local driving, for example:
	\begin{itemize}
		\item In the \textit{nonreciprocal XY model} \cite{LoosKM2023}, spins on a lattice align their orientations with that of their neighbors, but -- unlike in the standard XY model -- only with those neighbors that are in their vision cone. Since in general a spin 2 can be in the vision cone of a spin 1 without the converse being true, the interactions are nonreciprocal.
		\item In the \textit{nonreciprocal Ising model} \cite{AvniFMSV2023}, the lattice is occupied by two spin species A and B. Spins of species A have a tendency to align with those of species B, while spins of species B have a tendency to anti-align with those of species A.
	\end{itemize}
	In both of these cases, the system is found to exhibit TRS-breaking phases and thus steady-state informatic entropy production. However, in contrast to self-propelled particles or active nematics, and also in contrast to, e.g., the active Ising model \cite{SolonT2013}, the nonequilibrium nature does not give rise to translational motion of the individual constituents or a surrounding fluid. Motion is, at most, found at a many-body level in the form of traveling patterns (which can also arise in, e.g., reaction-diffusion systems that are not considered active). Thus, it is not completely clear whether these models qualify as active.
	
	A closely related issue arises in off-lattice models of systems with nonreciprocal interactions (for a more detailed discussion see Section \ref{nonreci} below). In this case motion and forces are usually present for the individual particles -- provided that there are several of them, since they do not in general move in isolation. Moreover, the directionality requirement holds in general only if one applies it to multi-particle configurations rather than to individual particles. Consider as an example the two particles shown at the top of \cref{fig:nri}: Since the red particle attracts the blue one and the blue one repels the red one, the two particles collectively move in a direction that depends on their configuration, and thus a dimer formed of both of them would comprise a self-propelled particle. However, each individual particle is symmetric and thus does by itself not satisfy the directionality requirement. The fact that activity is visible here only in collective phenomena was discussed very explicitly in the quote from Ref.\ \cite{SahaAG2020} cited in Section \ref{history}, and the classification of such systems as active may thus be thought of as an actual change of terminology (albeit a probably reasonable one), compared to what went before. Similar considerations apply for pulsating particles \cite{ZhangF2023}. The forces (or rather force fields) that emerge in a single-particle system are isotropic. However, if we consider a multi-particle system, the emerging pair-forces (and corresponding particle motions) have a direction.
	
	Another example comprises mixtures of particles driven by different thermostats \cite{GrosbergJ2018,GrosbergJ2015}. A practically relevant example for this comes from polymer physics \cite{SmrekK2017}: When DNA (a polymer) is transcribed, there are stronger fluctuations -- in effect, a higher temperature -- at certain positions of the polymer, namely at the points where transcription takes place. Considering the polymer to be a chain of beads, some of these beads would behave as if they were coupled to a different thermostat than their environment. This difference can, even if it is fairly small, give rise to phase separation. A related phenomenon is that of hot Brownian particles \cite{RingsSSCK2010,KroyC2023} that are maintained at a temperature above that of the solvent (such as laser-heated nanoparticles). Both systems are driven out of equilibrium by a local influx of energy, and thus they are active in a broad sense. However, they lack directional forces or motion that would result from a particle asymmetry, and are thus not always considered active in a narrow sense. 
	
	A final interesting borderline case comprises systems with time delay. A good illustration are Refs.\ \cite{KoppK2023,KoppK2023b,TaramaEL2019}, which have studied colloidal systems in which the force acting on a particle at time $t$ depends on its position at time $t-\tau$ with a delay $\tau$, and found that the dynamics of these particles resembles in many ways that of active Brownian particles. Theoretically, it can be found that such delay systems will generally have a non-vanishing entropy production \cite{MunakataIK2009,LoosK2019}. They are driven out of equilibrium by a local energy influx, this energy influx leads to motion. There can even be a directionality arising from particle properties, namely if the direction force acting on a particle depends on the vector separating the current and the previous particle position. However, studies of such delay models \cite{KoppK2023,KoppK2023b,TaramaEL2019} generally state that they find analogies to active matter rather than classifying the systems under study as themselves active.
	
The table in Fig.\ \ref{criteria} gives an overview of these criteria and whether they are satisfied by a number of systems discussed in this review.\footnote{Following Section \ref{howtofindit}, one could also think of \ZTP{viewed as active in the literature} as such a criterion.} This table illustrates that, in addition to a number of clear cases (such as birds and Janus particles), there are a number of systems where the classifcation is not fully obvious. All considered systems are out of equilibrium and have local energy influx, and this almost always leads to motion. Many systems generally though of as active show directionality. However, some non-directional systems (such as chemically driven phase-separating systems) are also considered active, and some directional systems (such as colloids with time delay) are not. 
	
	\section{\label{activefieldtheories}Active field theories}
	\textit{Being active} is something that is attributed both to physical objects (as in \ZTP{active particle} or \ZTP{active material}) and to mathematical theories (as in \ZTP{active field theory}). As already discussed in Section \ref{defining} above, the word \ZTP{active} does clearly not have the same meaning in both cases. So far this article has mostly focused on active particles. Here, we will address the meaning of \ZTP{active} interpreted as an attribute of a theory. To have a specific example, we will discuss the definition of the term \ZTP{active field theory}.
	
	Field theories are, in general, PDEs (either deterministic or stochastic) describing the stationary configuration or time evolution of a physical field $\phi(\vec{r},t)$ (or several physical fields, that can be scalar, vector or tensor in character), which depends on position $\vec{r}$ and time $t$ (in some cases also on other degrees of freedom such as orientation $\uu$). This field represents, for example, the local particle density or the local composition of a system. We are interested here in particular in the case where this field is conserved, i.e., where the integral of $\phi$ over space always has the same value. (If, for example, $\phi$ is the local particle density, then this correponds to particle number conservation.) This is done just to have a specific example, theories for non-conserved fields are also often found in active matter physics (a prominent example being active nematics \cite{DoostmohammadiIYS2018}).
	
	In the case of conserved fields and passive systems, deterministic field theories for diffusive motion typically have the form
	\begin{equation}
		\partial_t \phi = \Nabla\cdot\bigg(M\Nabla \Fdif{F}{\phi}\bigg),
		\label{deterministicch}
	\end{equation}
	with mobility $M(\phi)>0$ and free energy $F$. Important examples for this form are the Cahn-Hilliard equation \cite{Cahn1965,CahnH1958} and deterministic dynamical density functional theory (DDFT) \cite{Evans1979,MarconiT1999,ArcherE2004,teVrugtLW2020}. The stochastic counterpart of \cref{deterministicch} has the form
	\begin{equation}
		\partial_t \phi = \Nabla\cdot\bigg(M\Nabla\Fdif{F}{\phi}\bigg) + \Nabla \cdot(\sqrt{2Mk_\mathrm{B}T}\vec{\Lambda})
		\label{conservedch}
	\end{equation}
	with temperature $T$, and white noise $\vec{\Lambda}$ with zero mean and unit variance. This form\footnote{In principle, the use of \cref{conservedch} requires a specification of whether it is to be interpreted in the Ito or the Stratonovich sense. This, however, immaterial for the present discussion.} is in general referred to as \textit{Model B} \cite{HohenbergH1977}, important examples for this form are the Cahn-Hilliard-Cook equation \cite{Cook1970} and the Dean-Kawasaki equation \cite{Dean1996,Kawasaki1994,teVrugtLW2020}.
	
	It turns out that, if one starts from a microscopic description of an active particle system (specifically: overdamped active Brownian particles with repulsive interactions) and derives a field theory by coarse-graining (averaging), the result does not have the structure \eqref{conservedch}, but instead has the form \cite{TjhungNC2018,teVrugtBW2022}
	\begin{equation}
		\begin{split}
			\partial_t \phi &= \Nabla\cdot\bigg(M\bigg(\Nabla\Fdif{F}{\phi} + \lambda\Nabla(\Nabla\phi)^2 + \xi(\Nabla\phi)(\Nabla^2\phi)\bigg)\bigg)\\& + \Nabla \cdot(\sqrt{2Mk_\mathrm{B}T}\vec{\Lambda})
		\end{split}
		\label{amb}
	\end{equation}
	with parameters $\lambda$ and $\xi$ (that can also depend on $\phi$ in principle, though we suppress this dependence here). Equation \eqref{amb} is referred to as \textit{Active Model B+} and cannot be written in the form \eqref{conservedch} since the two terms involving $\lambda$ and $\xi$ cannot be written as (the Laplacian of) the functional derivative of any free energy $F$. The original motivation for the development of \cref{amb} \cite{TjhungNC2018} was to model motility-induced phase separation \cite{CatesT2015}, a widely studied collective phenomenon in active matter physics where active particles exhitit spontaneous phase separation into high- and low-density fluid phases without requiring attractive interactions (as they would in equilibrium). 
	
	Regardless of any specific microscopic derivation, there are systematic reasons why passive particles are described by equations of the form \eqref{conservedch}. Specifically, as discussed in Section \ref{modern}, the dynamics of particles in equilibrium obeys TRS, manifest in the fact that passive systems without an external drive have zero entropy production in steady state. The form \eqref{conservedch} of the field theory automatically ensures that this is the case, as it leads to a stationary probability distribution $P[\phi]\propto \exp(-\beta F)$, from which it is a simple exercise to prove that the entropy production \eqref{entropyproduction} vanishes in steady state \cite{Cates2019}. Turning this around: if the system is active, then it will usually have a non-zero entropy production in the steady state, and consequently it will be described by a field theory that does not prohibit such a non-zero entropy production and that therefore does not have the form \eqref{conservedch}. 
	
	Owing to these general considerations, it has become common to classify as active any theory for a conserved diffusive scalar  as active that breaks the structure \eqref{conservedch}, or, more precisely, that violates TRS and thereby produces entropy in the steady state \cite{Cates2019}. (The same criterion applies for non-conserved fields.) Such theories can then be constructed not only via a \textit{bottom-up approach} (coarse-graining a microscopic model that describes what the particles do), but more easily via a \textit{top-down approach} (writing down a theory of the form \eqref{conservedch} and then adding some TRS-violating terms, like the terms in $\lambda$ and $\xi$ in \cref{amb}). Indeed, active field theories like \cref{amb} have first evolved by writing down Landau-Ginzburg theory and empirically adding terms breaking detailed balance to a theory that has it \cite{Cates2019,WittkowskiTSAMC2014}.
	
	Theories of the form \eqref{amb}  have been developed for different systems, ranging from ones that are certainly active matter, such as self-propelled particles  \cite{TjhungNC2018}, via non-self-propelling particles with vision cone interactions \cite{HuangtVWL2024} and driven quantum systems \cite{SiebererBMD2023} through to models of social dynamics in human populations \cite{ZakineGBB2024}. The latter would not generally have been termed active matter prior to the use of Active Model B+ to describe them in Ref.\ \cite{ZakineGBB2024}. This is an example of how the meaning of \ZTP{active} can evolve over time. More generally, the study of active field theories thus often proceeds without commitment to any particular microscopic interpretation, and simply investigates in which ways a field theory that breaks TRS is different from a theory that does not. 
	
	As a consequence of this, there are in principle two ways of defining \ZTP{active field theory}, which are both problematic for different reasons:
	\begin{enumerate}
		\item An active field theory is a field theory intended to describe at a continuum level the dynamics of active particles. 
		\item An active field theory is a field theory that lacks a certain mathematical structure, namely a noisy gradient flow such as \cref{conservedch} involving a free energy $F$ and noise terms such that the stationary measure is $\exp(-\beta F)$ (with $\beta$, in general, some effective inverse temperature rather than a real one).  
	\end{enumerate}
	However, neither of these definitions is by itself fully satisfying.
	
	In the first case, one would essentially argue that, just as a quantum field theory is a theory for quantum fields, an active field theory is a theory for active fields. An active field would then be a field representing the state of an active matter system. A problem with this approach is that not every effect in active matter requires TRS breaking terms for describing it: sometimes active particles can (in an \textit{effective equilibrium framework}) be approximately modeled by an equation of the form \eqref{conservedch} \cite{EnculescuS2011,WittmannB2016}. Similarly, at a sufficiently high degree of coarse-graining, an ideal gas of active particles (clearly an active system) is described by the diffusion equation. Unless one wishes to classify the diffusion equation as an active field theory -- which would be strongly at odds with the way this term is is usually understood -- this kind of approach would obviously need some refinement. The problem is that if one simplifies or coarse-grains too far the activity can disappear altogether, and the resulting theory is not active even if it was intended as a model of active particles. (Specifically, if one ignores higher-order gradient contributions in the microscopic derivation of a model like \eqref{amb}, the nonvariational terms proportional to $\lambda$ and $\zeta$ disappear and one is left with an equation that has the structure of a passive theory \cite{CatesT2015}.) 
	
	The second option would be to consider \textit{active} to be a property of a field theory (that is, of a PDE or set of PDEs) with certain mathematical properties. This is similar in spirit to what is done in the theories of critical dynamics and pattern formation, where a classification for PDEs has been developed that relies on their mathematical properties rather than on their physical interpretation. One distinguishes, for instance, models with and without a conserved order parameter (referred to as \textit{Model B} and \textit{Model A}, respectively) \cite{HohenbergH1977}, or models with different types of pattern-forming instabilities \cite{CrossH1993}. This idea has been very fruitful, and has also influenced the field of active matter physics.
	
	A typical example for a characterization\footnote{This statement is not intended as a formal definition of \ZTP{active field theory} by the author of Ref.\ \cite{Cates2019}, but is nevertheless representative of the general understanding of this term in the literature} of \ZTP{active field theory} can be found in Ref.\ \cite{Cates2019}: \ZT{Active field theories involve, as equations of motion, stochastic PDEs for suitable order parameter fields [...] that are unconstrained by the requirements (i-v) above that all stem from microscopic TRS.}\footnote{The requirements (i-v) from TRS include the existence of a free energy functional, the Boltzmann distribution in steady state, the principle of detailed balance, the absence of steady-state currents, and the fluctuation-dissipation theorem}. The form \eqref{conservedch} ensures that all requirements of TRS are satisfied, and thus looking for deviations from this form is a promising way of identifying active field theories. Thus, it is customary to classify field theories as active if they have a non-zero entropy production in steady state \cite{NardiniFTvWTC2017}. This does, in practice, work well for field theories that have the form of stochastic PDEs, in other words, for field theories with a noise term. However, this sweepingly broad approach also has several problems. One of them is that it requires the system to have a state with a stationary measure, which may not be guaranteed. A more severe problem is that \textit{not having the time-reversal symmetry form \eqref{conservedch}} (or its generalization to other order parameters and conservation laws) is a requirement satisfied by a number of models that predate active matter physics and that one may not necessarily want to classify as active, such as reaction-diffusion models \cite{Turing1952}, field  theories for brain dynamics \cite{NartalloKDLG2025}, and the Kardar-Parisi-Zhang equation \cite{KardarPZ1986}. The latter was originally a model of irreversible interfacial growth but has been found to describe a wide range of other nonequilibrium systems, including some driven particle systems. Due to the coarse-grained nature of field theories, the distinction between active and driven systems, which is already problematic for particles, might become moot in the field-theoretical case, since a variety of active and driven systems may be coarse-grained to have the same continuum description. Moreover, the requirement \textit{field theory whose form is not constrained by the requirements of TRS} is satisfied by many arbitrary PDEs that have never been proposed as a description of any physical system, which would then suddenly count as active field theories. One can resolve the latter issue by arguing that these equations do, due to their lack of physical content, not constitute a theory (and therefore not an acitve field theory).
	
	A separate problem is that such definitions of \ZTP{active field theory} do not work for models like \cref{deterministicch}, which lack a noise term. In many cases, the absence of a noise term simply results from ignoring noise or from considering a limit (such as late stages of pattern formation processes \cite{CrossH1993}) where it does not matter. However, there are field theories whose derivation from a more microscopic model excludes the addition of a noise term (which, as noted in Ref.\ \cite{MarconiT1999}, does not always stop people from adding one when it is practically useful to do so). Such theories -- including many which are generally thought of as active, such as active dynamical density functional theories \cite{WensinkL2008,WittkowskiL2011,MenzelSHL2016}  or active phase field crystal models \cite{MenzelL2013,OphausGT2018,teVrugtJW2021,teVrugtHKWT2022,HuangML2020} describe the ensemble-averaged density and therefore do, in contrast to \cref{conservedch,amb}, not constitute \textit{statistical} field theories.\footnote{More precisely, these theories are obtained from the \textit{Smoluchowski equation}, which governs the time evolution of the probability distribution $P(\{\vec{r}_i,t\})$ that measures how likely it is that the particles can, at a time $t$, be found at the positions $\vec{r}_i$. While the Smoluchowski equation describes stochastic dynamics, it is not a stochastic equation since $P$ evolves deterministically. Thus, if we (as is frequently done) define the particle density as $\phi(\vec{r})=N\prod_{i}\TINT{}{}{r_i}\delta(\vec{r}-\vec{r}_i)P(\{\vec{r}_j,t\})$ and obtain the dynamic of equation for $\phi(\vec{r})$ by integrating out the one for $P(\{\vec{r}_j,t\})$, then the dynamic equation for $\phi$ cannot have a noise term \cite{ArcherR2004,teVrugtLW2020,teVrugt2020}. This definition of $\phi$ is a strict alternative to the one used in \cref{conservedch,amb} where $\phi$ is a fluctuating particle density undergoing stochastic evolution via a particle current whose mean value gives the deterministic term in the equation and whose variance sets the noise.} For deterministic theories, the definition \eqref{entropyproduction} of entropy production based on the likelihood of a given trajectory compared to the likelyhood if its time-reversal is not really applicable, since the probability of any trajectory is either zero or one.
	
	Still, one can often classify such theories as active or passive based on thermodynamic considerations. To see how, consider the multi-species version of \cref{deterministicch}, given by
	\begin{equation}
		\partial_t \phi_i = \Nabla\cdot\bigg(M_{ij}\Nabla \Fdif{F}{\phi_j}\bigg),
		\label{multispeciesch}
	\end{equation}
	where $M_{ij}$ is a mobility matrix that is positive definite and $\phi_i$ is the density of species $i$. (We sum over double indices.) From \cref{multispeciesch}, the time derivative of the free energy is easily found to be
	\begin{equation}
		\dot{F} = -\INT{}{}{^dr}M_{ij}\bigg(\Nabla \Fdif{F}{\phi_i}\bigg)\cdot\bigg(\Nabla \Fdif{F}{\phi_i}\bigg)  \leq 0,
		\label{dotf}
	\end{equation}
	with the number of spatial dimensions $d$, having used that the mobility matrix is positive definite. Thus, assuming for the moment that $F$ is bounded from below, the system evolves towards an equilibrium state in which one has $\dot{F}=0$ and in which the free energy $F$ is minimized (subject to the constraint that the overall mass is the same as in the initial state). An analogous argument can be made for the case that the dynamics has a non-conserved form (Allen-Cahn equation \cite{AllenC1979}), which in the deterministic multi-species case reads
	\begin{equation}
		\partial_t \phi_i = - C_{ij}\Fdif{F}{\phi_j},
		\label{multispecieschac}
	\end{equation}
	with a non-conserved mobility matrix $C_{ij}$ that is also positive definite. Equation \eqref{multispecieschac} implies
	\begin{equation}
		\dot{F} = -C_{ij}\INT{}{}{^3r}\Fdif{F}{\phi_i}\Fdif{F}{\phi_j}\leq 0.
		\label{dotf2}.
	\end{equation}
	Still, it should be noted that the important criterion here is not the fact that $\dot{F}=0$ in steady state but the fact that the model has the structure \eqref{multispeciesch}. A good way to see this is that Active Model B (AMB) \cite{WittkowskiTSAMC2014}, which is obtained from \cref{amb} by setting $\xi=0$, has no currents in steady state and therefore a constant $\phi$. Since the $F$ in \cref{multispeciesch} is a functional of $\phi$, we therefore need to have $\dot{F}=0$ in the steady state for AMB. Still, AMB has a non-vanishing entropy production in the steady state \cite{NardiniFTvWTC2017} and is therefore an active field theory. The deeper physical meaning of \cref{dotf} can be seen by noting that for a passive system with a steady current $\vec{J}$, the local entropy production is $\vec{J}^2/(Mk_\mathrm{B}T)$ \cite{Onsager1931} (with $T$ being the thermodynamic temperature). For a passive system, this is, up to prefactors, the integrand on the right-hand-side of \cref{dotf}, and in a noiseless system with stationary $\phi$, currents are the only way to get an entropy production.
	
	This suggests that a deterministic field theory can be considered passive simply if there are some $F$ and some $M_{ij}$ that allow it to be written in the form \eqref{multispeciesch}, and active otherwise. However, to infer from this the existence of an equilibrium state with $\dot{F}=0$ via an argument of the form \eqref{dotf} requires that $F$ is bounded from below (otherwise it may just continue to decrease) and that $M_{ij}$ is positive definite (otherwise the last inequality in \cref{dotf} does not hold). These criteria, which in equilibrium systems are generally taken for granted, can fail for active field theories. A good example to see this is the \textit{non-reciprocal Cahn Hilliard equation} \cite{SahaAG2020,YouBM2020,BraunsM2024}, which in its deterministic form reads
	\begin{align}
		\partial_t \phi_1 &= \Nabla^2\bigg(\Fdif{F}{\phi_1} - \alpha\phi_2\bigg),\label{nrc1}\\
		\partial_t \phi_2 &= \Nabla^2 \bigg(\Fdif{F}{\phi_2} +\alpha\phi_1 \bigg),\label{nrc2}
	\end{align}
	where $F$ is a free energy functional that we assume to have the form
	\begin{align}
		F[\phi_1,\phi_2] &= F_1[\phi_1] + F_2[\phi_2]+ F_{12}[\phi_1,\phi_2],\\
		F_{12}&=-\INT{}{}{^dr}\nu \phi_1\phi_2
	\end{align}
	and where $F_1$ and $F_2$ are free energy functionals depending solely on the fields $\phi_1$ and $\phi_2$, respectively. It constitutes a minimal field-theoretical model for multi-species non-reciprocal interactions (see Section \ref{nonreci}) as the two fields $\phi_1$ and $\phi_2$ are coupled non-reciprocally for $\alpha \neq 0$. The parameter $\nu$ encodes a possible reciprocal coupling.
	
	For the stochastic case, the non-reciprocal Cahn-Hilliard equation can be classified as as active on the basis that it has nonzero entropy production in steady state \cite{SuchanekKL2023}. In the deterministic case, however, it can be brought into the form \eqref{multispeciesch} using an argument from Refs.\ \cite{FrohoffHKGT2023,GreveFT2024}: If we define
	\begin{align}
		\tilde{F} &= \frac{\nu}{\nu + \alpha}F_1 + \frac{\nu}{\nu - \alpha} F_2 + F_{12},\\
		\underline{\tilde{M}} &=
		\begin{pmatrix}
			\frac{\nu+\alpha}{\nu} & 0 \\ 0 & \frac{\nu - \alpha}{\nu}
		\end{pmatrix}
	\end{align}
	with the amended free energy $\Tilde{F}$ and the amended mobility matrix $\underline{\tilde{M}}$. With these definitions, \cref{nrc1,nrc2} can be written as
	\begin{equation}
		\partial_t \phi_i = \Nabla\cdot\bigg(\tilde{M}_{ij}\Nabla\Fdif{\tilde{F}}{\phi_{j}}\bigg)\label{spuriuous}
	\end{equation}
	Equation \eqref{spuriuous} has exactly the same form as \cref{multispeciesch}, and is in Ref.\ \cite{GreveFT2024} used to develop phase diagrams. However -- and this is why \cref{spuriuous} is referred to as \ZT{spurious gradient dynamics} \cite{FrohoffHKGT2023,GreveFT2024} -- the mobility $\underline{\tilde{M}}$ is not positive definite and the free energy $\tilde{F}$ is not bounded from below. The physical consequence of this is that the system does not approach thermodynamic equilibrium and can, e.g., have a stationary measure within which the state of the system oscillates indefinitely. Crucially then, for a field theory to take the form \eqref{multispeciesch} is not sufficient to make it passive; the criteria on $F$ and $M$ referred to above must also be satisfied.
    
Moreover, also in the deterministic case one faces the problem that \textit{not having the structure \eqref{multispeciesch}} is a criterion satisfied by a large class of field theories that have no plausible utility in modeling active matter. Thus, both in the stochastic and the deterministic case, we currently lack satisfactory definition of \ZTP{active field theory}.
	
	\section{\label{coarsenature}The coarse-grained nature of activity}
	Regardless of which definition one decides to use, an important aspect that can be confusing for newcomers to the field is the fact that speaking about activity always corresponds to taking a coarse-grained point of view. Researchers who study active particles violating Newton's third law (i.e., particles with non-reciprocal interactions) do not actually claim or believe that there are physical systems for which the law of momentum conservation does not hold microscopically. Yet they choose a mesoscopic description that does violate Newton's third law. This happens because one ignores in the description environmental or internal states of the particles. For instance, an animal walking on the ground appears to not conserve momentum only because we do not explicitly take into account the momentum exchange with planet earth, and we choose not do do this because doing so would make the description way too complicated.\footnote{Similarly, the animal appears to not conserve energy only because we do not explicitly model its energy sources (the food it eats).} This implies that whether a system is active depends, to some extend, on how it is described, and that one can therefore not really speak about a physical system as \ZTP{active} independent of a theoretical model of it. On the other hand, the notion is also not completely arbitrary: a (sub)system can be thought as active if it has an autonomous dynamics that can be described with a closed stochastic equation that breaks detailed balance.
	
	Suppose, indeed, we were to model the entire universe in full atomistic (yet classical, i.e., non-quantum) detail. In this case, our description would be Hamiltonian and therefore passive. In practice though we will normally restrict our description to a few relevant variables. A paradigmatic example would be colloidal particles immersed in a fluid, where one considers only the positions $\vec{r}_i$ of the colloidal particles. We ignore in this case not only their internal degrees of freedom (by describing them simply as a hard spheres without explicitly considering its internal atomic structure) but also external degrees of freedom (the environment).  One can, via some coarse-graining procedure based, e.g., on projection operators \cite{Grabert1982,teVrugtW2019d,Schilling2022}, pass to a higher level of description involving only the colloid coordinates. This in general leads to equations with relatively complicated memory and noise terms \cite{teVrugtW2019d}. Under suitable conditions of time-scale separation and proximity to thermal equilibrium, we will however recover a Markovian dynamics that respects the principle of detailed balance, taking the form \cite{Zwanzig2001}
	\begin{equation}
		\dot{\vec{r}}_i = -\beta D \Nabla_{\vec{r}_i}U(\{\vec{r}_j\}) + \sqrt{2D}\vec{\xi}_i. 
		\label{langevin}
	\end{equation}
	(For a derivation of \cref{langevin} via projection operators see Ref.\ \cite{Zwanzig2001}, whose use of the term \ZTP{relevant variables} is adopted here.) What is notable about \cref{langevin} is that, even though we have ignored a substantial number of degrees of freedom, we can still find some potential $U$ from which all deterministic forces are derived and that tells us what the stationary probability distribution of the system is, namely $P(\{\vec{r}_j\}) \propto\exp(-\beta U(\{\vec{r}_j\}))$. Sometimes, however, it may turn out that the effective dynamics for the relevant variables have the form \eqref{langevin2} instead. In this case, we are still able to find useful equations describing the colloidal particles' degrees of freedom, but they no longer satisfy the requirement of detailed balance. This is what it means for a system to be active: There is an effective description (that may of course be more complicated than \cref{langevin2}, maybe even non-Markovian) in which we ignore the microscopic energy sources whose explicit inclusion would make the description passive (but still far from equilibrium until the fuel runs out). 
	
	An obvious question then concerns, say, a colloida particle in a gravitational field. If one ignores gravity as the source of energy, this may look like self propulsion (downward). Why not use such a description? In principle this would be perfectly fine. However, such a description would not be practically useful since in a reduced description that ignores the gravitational field, one would not have enough information to predict what the particle does. For the sorts of particles that are typically thought of as \ZTP{active}, however, the information about (for instance) position and orientation of the particle is sufficient to predict what it does. The actual energy sources are irrelevant.  Thus, the idea of a particle being active illustrates a general point about the relation of different levels of description of a physical system.
	
	This also helps to understand why systems with non-reciprocal interactions are often thought of as active. Fundamental interactions on the microscopic level are constrained by momentum conservation to obey Newton's third law. The idea of non-reciprocal interactions can nonetheless be useful for a reduced description. 
    
    To summarise the above, active theories represent effective models that operate at coarse-grained level only, and active field theories are likewise effective field theories which only aspire to be accurate at long wavelengths and timescales where the vast majority of microscopic degrees of freedom can be safely ignored.
	
As hinted at in Section \ref{whatitis} above, the coarse-grained nature of activity becomes important when we ask: What could be a criterion for determining whether a system is out of equilibrium? The entropy production seems a reasonable candidate. However, as discussed in Section \ref{modern}, one can, for the same system (the active particle governed by \cref{langevin}) find a number of entropy productions. On the one hand there is the informatic entropy production. Here, one can find three different results (\cref{epr1,epr2,epr3}) that depend on how one calculates it. Two of them (\cref{epr1,epr2}) give zero for a free self-propelled particle -- which might be taken to imply that it is not active. This is a manifestation of the fact that the notion of activity is to some extend dependent on the chosen level of coarse-graining. On the other hand, there is the thermodynamic entropy production. This is a quantity that \textit{for a given system} one can clearly identify and that thus gives a more objective idea of whether a physical system is out of equilibrium. Still, also the entropy of thermodynamics is to some extend an information-theoretical quantity (as Jaynes' \cite{Jaynes1992} discussion of the Gibbs paradox nicely shows), and also thermodynamic entropy production is in the end present because we use a coarse-grained description of the Hamiltonian microdynamics \cite{teVrugt2020}. 
	
	\subsection{\label{sec:janus}Example: Janus particles}
	A good way to appreciate the coarse-grained nature of activity is to consider scenarios in which one is able to explicitly derive the coarse-grained active dynamics from the fine-grained passive description. One such scenario was considered in Ref.\ \cite{Bebon2024}: A spherical colloidal particle with hemispheres of different surface chemistry (a Janus particle \cite{WaltherM2008}) is immersed in a solvent consisting of particles from two different species (substrate and product, denoted S and P) that interact with the colloid via different potentials $U_\mathrm{S}$ and $U_\mathrm{P}$, respectively. On one side of the colloid (which is catalytically active), the particles can undergo chemical reactions with rates $k^+$ (for S $\rightarrow$ P) and $k^-$ (for P $\rightarrow$ S) that satisfy the detailed balance condition
	\begin{equation}
		\frac{k^+(r,\theta)}{k^-(r,\theta)}= e^{-\beta (U_\mathrm{S}(\vec{r})-U_\mathrm{P}(\vec{r}))},
	\end{equation} 
	where the dependence of the reaction rates on the angle $\theta$ arises from the fact that the particles undergo reactions only on one side of the particle. This asymmetry defines the particle orientation, specified by an orientation vector $\uu$ (that in two spatial dimensions is a function of a single angle $\varphi$). Note that, so far, this setup corresponds to a passive particle.
	
	Next, the authors of Ref.\ \cite{Bebon2024} assumed that, at infinite distance from the colloidal particle, a chemostat keeps substrate and product particles at constant densities, which leads to a net chemical flux $\mathcal{J}$ between substrate and product reservoir. Each conversion of a substrate to a product particle pushes the particle a small distance $\lambda$ in the direction of its orientation (a product to substrate conversion has the same effect in the other direction). It can then be shown that the Smoluchowski equation for the probability $P(\vec{r},\varphi,t)$ of finding the particle at position $\vec{r}$ with orientation $\theta$ at time $t$ reads
	\begin{equation}
		\pdif{}{t}P = (- v_0 \Nabla\cdot \uu + D \Nabla^2 + D_\mathrm{R}\partial_\varphi^2)P,
		\label{smoluchowski}
	\end{equation} 
	where $v_0 = \lambda \mathcal{J}$ is the self-propulsion speed, $D$ and $D_\mathrm{R}$ are translational and rotational diffusion coefficients. Microscopic expressions for $\lambda$ and $\mathcal{J}$ can also be obtained. Notably, \cref{smoluchowski} is the Smoluchowski equation of an \textit{active} particle (an active Brownian particle \cite{RomanczukBELSG2012} to be precise). Consequently, the combination of chemostatting and ignoring the solvent dynamics has led from a passive to an active model. This example is particularly helpful because it illustrates the characteristic property of active matter: it exchanges energy and momentum with some larger environment, but has an autononmous sub-dynamics that on its own would violate general principles of Hamiltonian mechanics and/or of thermodynamics.
	
	\subsection{Example: Reactive droplets}
	For the field-theoretical case, a very similar result was obtained in Ref.\ \cite{VossT2024} in a model for reactive thin-film hydrodynamics. The authors considered a droplet, described by a film height $h(\vec{r},t)$ (that depends on position $\vec{r}$ -- a two-dimensional vector giving the position in the $xy$ plane -- and time $t$), on a substrate. Particles with height-averaged density $a(\vec{r},t)$ are suspended in the droplet. They can also adsorb to the substrate; the density of adsorbed particles is given by $b(\vec{r},t)$. These quantities evolve according to the dynamic equations (cf. Eqs.\ (45)-(47) in Ref.\ \cite{VossT2024})
	\begin{align}
		\partial_t h&= \Nabla \cdot \bigg(Q_{hh}\Nabla\Fdif{F}{h}+Q_{ha}\Nabla\Fdif{F}{a}+Q_{hb}\Nabla\Fdif{F}{b}\bigg) \label{partialth},\\
		\partial_t a&= \Nabla \cdot \bigg(Q_{ah}\Nabla\Fdif{F}{h}+Q_{aa}\Nabla\Fdif{F}{a}+Q_{ab}\Nabla\Fdif{F}{b}\bigg) + J_r\label{partialta},\\
		\partial_t b&= \Nabla \cdot \bigg(Q_{bh}\Nabla\Fdif{F}{h}+Q_{ba}\Nabla\Fdif{F}{a}+Q_{bb}\Nabla\Fdif{F}{b}\bigg) - J_r\label{partialtb}
	\end{align}
	with the free energy $F$ and the rate $J_r$ at which particles of $a$ and $b$ type can be converted into each other (which physically corresponds to adsorption at or desorption from the substrate). This is, by the classification proposed in Section \ref{activefieldtheories}, a passive field theory. While at some point the fuel ($a$) will run out and the system will go to equilibrium, in practice the number of particles within the droplet is so large that there is a quasi-stationary regime before this happens in which $\partial_t a=0$. With this assumption, \cref{partialth,partialtb} simplify to (cf Eqs.\ (51)-(52) in Ref.\ \cite{VossT2024})
	\begin{align}
		\partial_t h&= \Nabla \cdot \bigg(Q_{hh}\Nabla\Fdif{F}{h}+Q_{hb}\Nabla\Fdif{F}{b}\bigg)\label{partialtha},\\
		\partial_t b&= \Nabla \cdot \bigg(Q_{bh}\Nabla\Fdif{F}{h}+Q_{bb}\Nabla\Fdif{F}{b}\bigg) - J_r\label{partialtba}
	\end{align}
	The authors of Ref.\ \cite{VossT2024} argue that \cref{partialtha,partialtba} constitute, except for special chemostatting choices, an \textit{active} field theory since the system does not relax to a state in which all currents separately vanish (i.e., to equilibrium). A numerical solution of \cref{partialtha,partialtba} shows that they describe self-propelling droplet. This case illustrates for fields what Section \ref{sec:janus} has shown for particle-based models: An active model (\cref{partialtha,partialtba}) arises from a passive one (\cref{partialta,partialtb,partialth}) by moving to a more coarse-grained level of description, within which the mechanism for sustaining the system away from equilibrium is no longer explicit.
	
	\subsection{Example: Ultrasound-driven particles}
	In the field of artificial active matter, much recent work has been directed at ultrasound-driven propulsion. Here, aspherical particles in an ultrasound field experience forces resulting from a momentum transfer from acoustic waves to the particle, as a consequence of which the particle moves. Experimentally, this mechanism was demonstrated by \citet{WangCHM2012}, and a theoretical model for this experiment was proposed by \citet{NadalL2014}. A key advantage of ultrasonic propulsion in the context of medical applications like drug delivery is that of biocompatibility -- no potentially toxic chemicals are required to achieve propulsion \cite{LuoFWG2018}. Further benefits of this approach are tunability and high propulsion speeds \cite{McNeillM2023}. An important mechanism for the propulsion of ultrasound-driven particles is acoustic streaming \cite{NadalL2014}. Here, an acoustic wave in a viscous fluid leads to a non-oscillatory flow (a second-order effect that is present in addition to the oscillatory fluid motion) \cite{Wu2018}. In addition, radiation forces can be relevant \cite{VossW2020}. Reviews can be found in Refs. \cite{McNeillM2023,LiMOP2022}.
	
	One may at first not be inclined to call an ultrasound-driven particle \ZT{active}. Unlike a bacterium, it is powered not by internal fuel, but pushed by an external field, and there is an asymmetry in the propulsion that results from the external ultrasound wave rather than from the particle's asymmetry. In fact, many authors in this field do not even employ the term \ZTP{active particles} for the objects they study, more common are formulations like \ZT{acoustically actuated microrobot} \cite{DengPZWA2023}. However, as emphasized above, being active is not an observer-independent feature of a particle, but an aspect of the way it is described. While a microscopic description as used by \citet{VossW2020} is passive as it is simply based on solving the Navier-Stokes equation for a fluid forced by an incoming ultrasound wave, one can employ an effective description in which one ignores the ultrasound (and the surrounding fluid in general) and simply views the particle as having a self-propulsion force whose direction (and in this case also magnitude) depends on the particle orientation. On this level of description, the particle is an active object, which exhibits standard active matter phenomena such as motility-induced phase separation \cite{BrokerBtVCW2023}. In particular, the particle does, on this level of description, satisfy the directionality criterion since the propulsion mechanism relies on the asymmetric particle shape \cite{NadalL2014}. Spherical bubbles simply oscillate in traveling ultrasound waves, while particles with asymmetric shape \cite{NadalL2014,McNeillM2023} (or mass density \cite{NadalM2020}) experience directional motion \cite{McNeillM2023}. Moreover, simulations show that the hydrodynamic field near an ultrasound-driven particle is similar to that of a certain type of active particle (specifically, a so-called pusher squirmer) \cite{VossW2020}.
	
	In summary, the role of the acoustic field in ultrasonically driven particles is similar to that of the vibrational forcing in a granular layer \cite{NarayanRM2007}, as discussed in Section \ref{nemat} below (with the exception that the former system is wet, the latter dry). In both cases, the actual direction of motion is set by the internal coordinates (specifically orientation) of anisotropic particles, not by the external forcing itself. This is one reason why it is fruitful to consider both systems as comprising active particles.
	
	\section{Recent developments in active matter physics}
	We will now illustrate some aspects discussed in previous sections with reference to various types of active matter -- namely active nematics, systems with non-reciprocal interactions, and quantum active matter -- that have been a focus of research in the past years. We will illustrate in which sense these systems can be though of as active matter even though they are, unlike the examples from the prevous section, not (in general) self-propelled at the particle level. In passing, we will thereby also provide a review of some recent developments in the field of active matter physics.
	
	\subsection{\label{nemat}Active nematics}
	Active nematics were among the first systems studied in active matter physics \cite{MishraR2006,ChateGM2006,AditiR2002} and continue to be a widely investigated material class \cite{RozmanY2024,SchimmingRR2024,AlamNSLGYBFD2024,CoelhoFT2023}\footnote{... which justifies discussing them in a section entitled \ZT{Recent developments in active matter physics}.}. Despite this the term \ZTP{active nematic} is elusive to define; the difficulties exemplify the  definitional question for active matter more generally.
	
	It is helpful to start with what \ZTP{nematic} means. Most nematic liquid crystals \cite{Friedel1922} consist of elongated (rod-like) particles that align their long axes with each other \cite{Stark2001}. In many cases these particles have a head-tail-symmetry (apolar rods). In the case of self-propelling active particles there has to be some asymmetry that determines the direction of self-propulsion, and therefore active particles with a head-tail-symmetry are generally not self-propelled.\footnote{A nematic phase is equally possible in systems without head-tail-symmetry so long as the particles point up and down the preferred axis with equal probability. Adding self propulsion in this case also generates an active nematic \cite{DombrowskiCCGK2004,CisnerosCDGK2007}.} They can, nevertheless, generate some other kind of motion systematically. Important examples are \ZT{shakers} \cite{BaskaranM2012,PutzigB2014}, which, as the name suggests, undergo nondirected shaking motion due to local energy influx. An experimental realization of shakers are vibrated apolar granular rods \cite{NarayanRM2007}. 
	
	A key feature of non-propelling active nematics that unifies them with self-propelled particles is that in both cases ordering can give rise to density currents \cite{BaskaranM2012}. In the case of self-propelled particles, whose ordering is represented by the polarization $\vec{P}$ (direction of the self-propulsion force), the current in a dry model has a contribution proportional to $\vec{P}$ \cite{MenzelL2013}. For nematics, whose ordering is represented by the nematic tensor $Q$, the lowest-order vector one can construct is $\Nabla\cdot Q$. TRS forbids a contribution $\Nabla\cdot Q$ to the current. Active nematics, however, are not constrained by TRS, and thus a field theory for active nematics often has a term $\Nabla\cdot Q$ in the current (in the case of a dry model) or the force density (in the case of a wet model).
	
	Since active nematics include systems that at individual particle level have head-tail symmetry and hence are incapable of self-propulsion, they create an important obstacle to \cref{defd,deff} according to which active nematics are not examples of active matter. Instead their activity stems from forces they exert on thei surroundings (wet) and/or particle currents emerging at collective rather than individual level.  Moreover, active nematics provide a case study for how self-propulsion at a coarse-grained level can emerge from a general nonequilibrium dynamics at a more microscopic level: It has been found that topological defects (specifically, those with $+1/2$ charge) in active nematics often exhibit directed motion and can thus be modeled as self-propelled particles \cite{GiomiBMSC2014}. 
	
	\subsection{\label{nonreci}Non-reciprocal systems}
\begin{figure}
		\centering
		\includegraphics[scale=0.3]{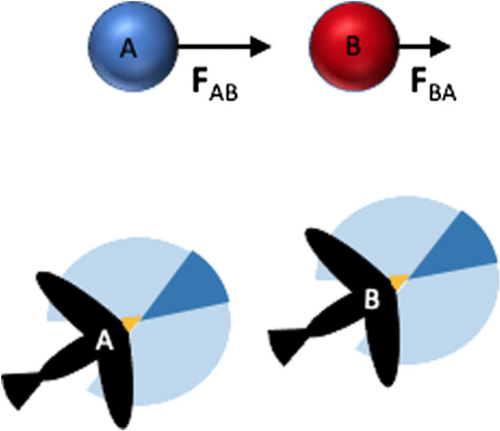}
		\caption{Particles with nonreciprocal interactions. (Reproduced with permission from Ref.\ \cite{BowickFMR2022}.)}
		\label{fig:nri}
	\end{figure}

	An important class of systems that have been widely studied by active matter physicists are those with non-reciprocal interactions. The typical example for this are particles violating Newton's third law (see Section \ref{modern} and \cref{fig:nri}). Non-reciprocal interactions are fairly common -- in particular in biological systems they are probably the rule rather than the exception \cite{HickeyGV2023,OuazanAG2023,VosseldGG2023}. Thus, while non-reciprocal interactions have come to the fore in active matter physics only rather recently, they have been studied in other fields for many years. Predator-prey models were developed already at the beginning of the 20th century \cite{Lotka1925}. The fact that thermodynamic reciprocity relations can be violated by systems that break TRS on a microscopic level was already noted by \citet{Onsager1931,Onsager1931b}. Biological and artificial neural networks -- a paradigmatic example for systems with non-reciprocal interactions -- have been studied for decades \cite{Rosenblatt1958,Kabrisky1966,Fukushima1980}. And finally, systems of spins with non-reciprocal interactions were already investigated in the 1980s, then known as \ZT{spins on directed graphs} \cite{ChowW1987}.
	
	Notably, while there is almost universal agreement that non-self-propelling active nematics are active, systems with non-reciprocal interactions are more of a borderline case (and in some cases obviously passive, see below). 
	Broadly speaking, one can distinguish two types of non-reciprocal interactions \cite{HuangtVWL2024}. In \textit{single-species nonreciprocity}, there is one particle species within which the particles interact non-reciprocally. A paradigmatic example, visualized in \cref{fig:nri} (bottom), are particles with a vision cone \cite{BarberisP2016,NegiWG2022,LoosKM2023,HuangtVWL2024}. These experience forces only from particles they can see, which depends on whether these are within their vision cone angle (shown in dark blue). In this case, particle A can see can see particle B but not vice versa, and only one of them will experience a force. The system thus breaks Newton's third law. In \textit{multi-species nonreciprocity} particles of one species A react different to particles of another species B than particles of species B react to particles of species A (see \cref{fig:nri}, top) -- or, more precisely: The forces that A particles exert on B particles and the forces that B particles exert on A particles cannot be derived from the same Hamiltonian. Here, a paradigmatic example are predator-prey systems \cite{ChenK2014,MeredithMGCKvBZ2020,teVrugtJW2024}, where the predator is attracted by the prey but the prey is repelled by the predator. Further examples for multi-species nonreciprocity can be found in mixtures of active and passive particles \cite{StenhammarWMC2015,WittkowskiSC2017,YouBM2020} or active molecules made of isotropic building blocks \cite{MandalSA2024}.
	
	It has become quite common to refer to particles with non-reciprocal interactions as \ZTP{active} \cite{SahaAG2020,HuangtVWL2024}. One motivation behind this is that particles with non-reciprocal interaction share some similarities with self-propelled particles. They often exhibit spontaneous motion (which non-reciprocal interactions can give rise to even in the absence of self-propulsion forces) in the absence of external potentials, and they can break TRS and thereby continuously produce entropy. In fact, many of these systems do thereby satisfy many or even all of the criteria for activity discussed in Section \ref{whatitis}.
	
	So far we have been concerned with systems that are non-reciprocal in the sense that they break Newton's third law. Physics knows many other reciprocity relations of similar type -- an important example is the \textit{Maxwell-Betti reciprocity theorem} \cite{Maxwell1864,Betti1872}, which states that in an elastic medium the displacement which a force applied at point A causes at point B is equal to the displacement which a force applied at point B causes at point A \cite{CoulaisSA2017}. A theorem by \citet{Onsager1931} and \citet{Casimir1945} states that reciprocity relations of this form generally hold in a linear time-invariant medium which is close to thermodynamic equilibrium provided that the microscopic dynamics of the system satisfies TRS \cite{NaessEHK2009}. Conversely, if the microscopic dynamics of the system does not obey TRS, then such reciprocity relations can be broken, which in the case of a violated Maxwell-Betti relation gives rise to a phenomenon \textit{odd elasticity} \cite{ScheibnerSBSIV2020,FruchartSV2023}. Here, the elastic modulus tensor $C_{ijmn}$ appearing in Hooke's law $\sigma_{ij}=C_{ijmn}\partial_m u_n$ with the stress tensor $\sigma_{ij}$ and the displacement vector $u_n$ has an odd component with $C^\mathrm{o}_{ijmn} = - C^\mathrm{o}_{mnij}$ \cite{ScheibnerSBSIV2020}. (This is possible also in momentum-conserving systems \cite{ScheibnerSBSIV2020}.) Materials exhibiting properties such as odd elasticity are widely referred to as active \cite{ScheibnerSBSIV2020}, even though their physics can be quite different from that of the self-propelling microswimmers or birds that one usually thinks of when hearing the term \ZTP{active matter}. The motivation for this development are ideas encoded in \cref{defe} -- the constituents of a material breaking thermodynamic reciprocity relations should have a TRS-violating dynamics, and in this sense, corresponding systems can be thought of as active.
	
	However, there are also examples for systems with nonreciprocal interactions where there is widespread agreement that they are \textit{not} active. A good example are colloidal particles with hydrodynamic interactions. These arise because the motion of colloidal particles through a fluid induces a flow field which in turn affects the motion of other particles moving through this fluid. Instead of solving the equations of motion for particles and fluid motion explicitly, one usually writes down equations motion for the particles. These contain hydrodynamic interaction forces that depend on the positions and velocities of the other particles \cite{Dhont1996}. Thus, also hydrodynamic interaction forces arise by switching to a coarse-grained point of view. Moreover, hydrodynamic interactions in a nonequilibrium (e.g., sedimenting) system can be nonreciprocal \cite{PoncetB2022,GuilletPLBIVB2025}. And finally, depending on how one interprets the local energy influx requirement, one could argue that it is also satisfied here: When the particles induce a fluid flow, they lose energy to the fluid, and they gain energy from the fluid when they are transported by a flow induced by the motion of other particles. Thus, when looking only at the particles, one sees a continuous influx and dissipation of energy. (Note that this holds in steady state only when the overall system is being driven.) This nonequilibrium nature is also manifest in the dynamics, for example in the fact that sedimenting particles with hydrodynamic interactions exhibit time-periodic behavior in steady state \cite{CaflischLLS1988}.  
	
	Nevertheless, there is no literature we are aware of were it is claimed that particles with hydrodynamic interactions should be classified, in general, as active. In fact, \cref{defe} explicitly excludes the scenario of sedimenting particles, with an appeal to what we have in Section \ref{whatitis} termed the directionality requirement. However, a directionality requirement is not employed in all definitions (it is absent in \cref{defb,defc}). Moreover, as discussed in Section \ref{whatitis}, there are a variety of systems (such as the symmetric active colloids discussed in Refs.\ \cite{Golestanian2009,ZhangF2023}) which are often considered active that do not satisfy the directionality criterion. However, without the directionality criterion, it is quite difficult to give a precise argument for why sedimenting colloids with hydrodynamic interactions should not be active. This issue is thus a case study for the problems addressed in Section \ref{howtofindit}: Depending on how we define \ZTP{active matter} -- with or without the directionality requirement -- we end up with a definition that either excludes systems that we do not want to exclude (symmetric active colloids) or that includes systems that we do not want to include (sedimenting particles with hydrodynamic interactions). While in the long-studied case of particles with hydrodynamic interactions there is a general agreement on their status (not active) even in the absence of such a criterion, this issue gains more practical relevance if one identifies new classes of systems with nonreciprocal interactions and asks whether or not they are active.

	\subsection{Liquid-liquid phase separation in biology}
	The function of biological cells relies heavily on their internal structure (\textit{compartmentalization}). While this is often achieved via membranes, there are also a number \textit{membraneless organelles}, such as centrosomes and the nucleolus, which form despite the absence of a membrane through liquid-liquid phase separation (LLPS) and behave like liquid droplets
	\cite{ShinB2017,HymanWJ2014,SchaferS2025,JulicherW2024}. LLPS in biology is a rapidly growing field of research. This has a variety of reasons, including the fact that it provides interesting applications and research directions for the physics of phase separation\footnote{\citet{McLeish2020} has provided an apt description of this development: \ZT{There is a rather serious collective phenomenon going on at the moment in the physics of liquid-liquid phase separation (LLPS) in the form of a widespread excitement about its application to cell biology.}} and the importance of LLPS for biological processes such as diseases and aging \cite{DormannL2024}. While equilibrium thermodynamics is very successful in explaining many aspects of this process, the cell is driven out of equilibrium by chemical reactions, and by other processes involving, e.g., molecular motors and ion pumps. This may play an important role for the physics of phase separation \cite{CatesN2025}, as has been shown both on a macroscopic \cite{WeberZJL2019,JulicherW2024} and a microscopic \cite{ZippoDSS2025} level. In general, both passive and active phase separation mechanisms may be relevant in this context \cite{BerryBH2018}, presumably to different extents in different systems.
	
	Making a link between intracellular phase separation and active matter physics is also plausible on phenomenological grounds. For instance, it is known that active phase separation can lead to arrested coarsening and the formation of bubbles in droplets \cite{TjhungNC2018}, and intracellular condensates do show arrested coarsening \cite{JulicherW2024} and bubble formation \cite{KistlerEtAl2018,ModiNLLBO2025}. However, some of these features have also been reproduced in in-vitro-experiments without any sort of energy influx \cite{ErkampEtAl2023}. In practice, the question of whether and when LLPS in cells should be viewed an active process may depend on the system under consideration \cite{McLeish2020}. For instance, Refs.\ \cite{ErkampEtAl2023,SalehJL2020,SalehWSL2023} all report experimental observations of bubbly phase separation. In Ref.\ \cite{ErkampEtAl2023}, this is attributed to a kinetic arrest (passive), whereas in Refs.\ \cite{SalehJL2020,SalehWSL2023} it is attributed to enzymatic degradation (active). For the case of arrested coarsening, \citet{McLeish2020} explicitly contrasts several works reporting different possible mechanisms (namely hydrodynamics \cite{SinghC2019} and chemical reactions \cite{WeberZJL2019}), and also discusses that this does not necessarily have to be an active effect \cite{SweatmanL2019}.
	
	From a terminological perspective, chemically driven LLPS is an interesting case since in this field it is quite common to refer to it as an active process \cite{JulicherW2024,SamantaBMW2024,WeberZJL2019} -- despite the fact that the phenomena here are (with some exceptions \cite{SalehWSL2023,TayarCASCD2023}) rather distinct from those studied in traditional active matter physics. Indeed, LLPS of this type is similar to nonequilibrium reaction-diffusion systems that have been studied long before the advent of modern active matter physics. The fact that the term \ZTP{active} is still frequently used in the study of LLPS may be a consequence of overall increase of popularity among physicists. Still, as the aforementioned connections between field theories for motility-induced phase separation \cite{TjhungNC2018} (traditional active matter physics) and field theories for LLPS \cite{WeberZJL2019,JulicherW2024} shows, this shared terminology can be very valuable for gaining physical insights.
	
	\subsection{\label{aqm}Quantum active matter}
	The vast majority of systems studied in active matter physics are sufficiently large, slow, and warm that one can safely treat them as being entirely classical. This, however, is beginning to shift. Already for some time, it has been common to study (mathematical) analogies between quantum and active systems \cite{LopezGDAC2015,MengMMG2021,LoeweSG2018,teVrugtFHHTW2023}. Particularly prominent in this context are links made between quantum systems with non-Hermitian Hamiltonians and non-Hermitian band theory in classical active systems \cite{ShankarSBMV2020}. Recently, there have been attempts to go beyond analogies and to realize genuinely quantum-mechanical systems that exhibit features of activity \cite{YamagishiHIO2023,TakasanAK2023,ZhengL2023,YamagishiHO2023,KhassehWMWH2023}. These developments illustrate the importance of finding a precise characterization of the term \ZTP{active matter}. Since quantum systems are quite different from the systems usually considered by active matter physicists, one is here less likely to have an intuitive idea about which systems should or should not be called \ZTP{active}. Instead, one needs to formulate some criteria that an active system needs to satisfy, and then try to identify a quantum system that meets these criteria.
	
	On the one hand, since intrinsically self-propelled quantum particles are not yet available, one might at first glance be sceptical whether such a thing as quantum active matter can exist at all (except in the trivial sense that all physical objects, including active ones, are fundamentally quantum). It is, for many forms of active matter typically considered, quite difficult to imagine how one might realize them on the level of individual atoms or molecules. In particular, classical active particles are typically immersed in a medium -- for example a fluid -- that they exchange momentum with, and this is crucial for them being able to self-propel. Such a surrounding fluid, which in general can also exchange heat energy with the particles, is not present in quantum setups like the one studied in Ref.\ \cite{ZhengL2023}. Another issue is that active particles, in order to generate systematic motion from free energy, need to have a sufficient amount of internal structure \cite{Ramaswamy2017}, and one may be sceptical as to whether small quantum systems are complex enough to possess this kind of structure. (The latter argument is not very strong though since also in the classical case there is often not that much structure required to generate systematic motion \cite{Ramaswamy2017}, and recent studies indicate that one may be able to design quantum systems with this kind of structure \cite{PennerEtAl2025}.) Perhaps to meet this sort of scepticism, authors in this field tend to be very careful in their wording and usually do not claim to realize quantum active matter, instead they speak about quantum analogies \cite{TakasanAK2023} or quantum mimicking \cite{ZhengL2023} of activity or self-propulsion. (However, \citet{NadolnyBB2025}, for example, do explicitly claim to present a realization of quantum active matter.)
	
	Nevertheless, at least for broader definitions of \ZTP{active matter} it is difficult to see why a very small and genuinely quantum-mechanical system should not be able to satisfy them. For instance, there have been a number of experimental and theoretical studies of quantum systems with non-reciprocal interactions \cite{RieserEtAl2022,ReisenbauerEtAl2024,LivskaEtAl2024,RudolphDHS2024,NadolnyBB2025}. If one takes classical particles with non-reciprocal interactions to be active -- which, as discussed in Section \ref{nonreci}, many people do -- it is hard to see why one should not do the same for their quantum counterparts. Thus, terminological differences here are probably just imported from the classical case -- for instance, both \citet{RieserEtAl2022} and \citet{NadolnyBB2025} present quantum systems with nonreciprocal interactions, but only \citet{NadolnyBB2025} refer to them as \ZT{active}. Moverover, there have been several studies \cite{delSerL2023,HardtDDdSR2025} showing that magnetic skyrmions and magnetic domain walls can exhibit self-propulsion. \citet{HardtDDdSR2025} suggest this as a possible route for the realization of quantum active matter.
	
	The operating principle in many studies of nonreciprocal interactions in quantum systems is based on light-matter couplings, which lead to nonreciprocal interactions so long as the coupling to the electromagnetic field is not explicitly considered (coarse-grained description). For example, in the experiments of Ref.\ \cite{RieserEtAl2022}, two nanoparticles are located in two distinct optical traps. They experience forces both from their optical trap and from the light scattered from the other particle. The resulting non-reciprocality leads to non-Hermitian dynamics involving parity-time symmetry breaking \cite{ReisenbauerEtAl2024} (which is typical for non-reciprocal systems \cite{FruchartLV2021}). Similar setups were previously used in the classical case to generate nonreciprocal interactions \cite{DuanBLD2022}.
	
	Too broad a definition of activity, such as one primarily requiring local energy influx, has the consequence that a very large number of systems that have long been studied in the physics of open quantum systems \cite{SiebererBMD2023} would be classified as active. A laser, for instance, is driven out of equilibrium by local influx of energy (and \ZTP{active medium} is indeed a common technical term in this context). \citet{ReisenbauerEtAl2024} do, in fact, refer to the non-reciprocity-induced oscillatory dynamics they observe as a \ZT{mechanical lasing transition}. Note, however, that it is not in principle a problem if a new understanding of \ZTP{quantum active matter} covers some or even many previously known systems. Active matter physicists are surely not the first people who study bird flocks, yet they have developed a new and fruitful way of looking at them, namely by viewing them as condensed matter systems \cite{CavagnaG2014}. Similarly, classifying systems that have been studied under the name \ZTP{driven quantum system} as \ZTP{quantum active matter} and investigating them from this point of view may be a very valuable enterprise, particularly if it allows to link the soft matter and the quantum physics communities.
	
	We now discuss two specific proposals for realizing quantum-mechanical active matter. The first one, developed in Refs.\ \cite{AdachiTK2022,TakasanAK2023}, is a lattice model that can be thought of as a quantum extension of the active Ising model \cite{SolonT2013} and is reported to exhibit (quantum analogues of) motility-induced phase separation and flocking transitions \cite{AdachiTK2022}. Similar ideas were employed also by other authors \cite{KhassehWMWH2023}. Specifically, one has a one-dimensional lattice with $L$ sites in which each site is either empty or occupied by one particle (not more than one, as the particles are assumed to have hard-core interactions) that can have spin $s=\pm 1$. Operators $\hat{a}_{i,s}^\dagger$ and $\hat{a}_{i,s}$ create and annihilate a particle with spin $s$ at site $i$. Based on them, one can define a particle number operator $\hat{n}_{i,s}= \hat{a}_{i,s}^\dagger\hat{a}_{i,s}$ as well as spin operators $\hat{m}_i^z = \hat{n}_{i,1}-\hat{{n}}_{i,-1}$ and $\hat{m}_i^x = \hat{a}_{i,+1}^\dagger \hat{a}_{i,-1} + \hat{a}_{i,-1}^\dagger \hat{a}_{i,+1}$. Using these operators, we construct the non-Hermitian Hamiltonian
	\begin{align}
		\hat{H} & = \hat{H}_{\mathrm{hop}} + \hat{H}_{\mathrm{act}} + \hat{H}_{\mathrm{TFIM}}\label{activehamil},\\
		\hat{H}_{\mathrm{hop}}&= - t \sum_{i=1}^{L}\sum_{s=\pm 1}(\hat{a}^\dagger_{i+1,s}\hat{a}_{i,s}+\hat{a}^\dagger_{i,s}\hat{a}_{i+1,s}),\label{hhop}\\
		\hat{H}_{\mathrm{act}}&= - \epsilon t \sum_{i=1}^{L}\sum_{s=\pm 1}s(\hat{a}^\dagger_{i+1,s}\hat{a}_{i,s}-\hat{a}^\dagger_{i,s}\hat{a}_{i+1,s}),\label{hact}\\
		\hat{H}_{\mathrm{TFIM}}&= - J_z \sum_{i=1}^{L}\hat{m}_i^z \hat{m}_{i+1}^z - h_x \sum_{i=1}^{L}\hat{m}_i^x\label{htfim}
	\end{align}
	with positive parameters $t$, $\epsilon$, $J_z$, and $h_x$.
	
\begin{figure}
		\centering
		\includegraphics[scale=0.6]{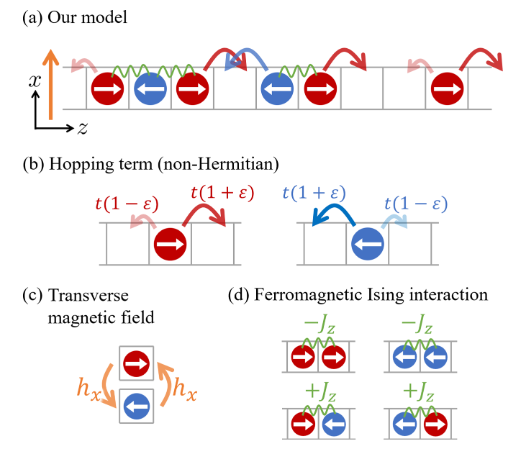}
		\caption{Visualization of the model defined by the Hamiltonian \eqref{activehamil} describing a quantum-mechanical active system. The figures show (a) the dynamics of the model, (b) hopping induced by $\hat{H}_{\mathrm{hop}}$ (\cref{hhop}) and $\hat{H}_{\mathrm{act}}$ (\cref{hact}), effects of the transverse magnetic field ($\hat{H}_{\mathrm{TFIM}}$, first term in \cref{htfim}), and (d) magnetic interactions ($\hat{H}_{\mathrm{TFIM}}$, second term in \cref{htfim})(Reproduced with permission from Ref.\ \cite{TakasanAK2023}.)}
		\label{fig:aqm1}
	\end{figure}

	Figure \ref{fig:aqm1} visualizes the physical meaning of this Hamiltonian. It describes particles hopping around on a one-dimensional lattice with a preferred direction that depends on their spin (\cref{fig:aqm1}a). This hopping (\cref{fig:aqm1}b) is encoded in the hopping terms $\hat{H}_{\mathrm{hop}}$ and $\hat{H}_{\mathrm{act}}$. The Hermitian operator $\hat{H}_{\mathrm{hop}}$ is widely used in condensed matter physics to describe how spins move (hop) from one atom to another one . The active term $\hat{H}_{\mathrm{act}}$ has a similar structure, but, crucially, with a relative sign that depends on the spin $s$. As a consequence, particles with spin $s=+1$ move to the right with a rate $t(1+\epsilon)$ and to the left with a rate $t(1-\epsilon)$, 
	Intuitively, it is reasonable to think of such a system as active. The particles have an orientation (their spin) and tend to move in the direction of their orientation. The operator encoding this motion is non-Hermitian. Finally, the Hamiltonian $\hat{H}_{\mathrm{TFIM}}$ of the transverse-field Ising model describes particles in a magnetic field pointing in the $x$ direction (orthogonal to the chain, \cref{fig:aqm1}c) that can change their direction owing to ferromagnetic (alignment) interactions (\cref{fig:aqm1}d).
	
	One may now be wondering why a system should have such a Hamiltonian. In this regard, this quantum active matter model is helpful as an example to see how activity emerges because the system is actually a subsystem of a larger system whose remaining degrees of freedom can be eliminated from the description. The dynamics of a quantum system with density operator $\hat{\rho}$ that is in contact with some environment is often described by the Lindblad equation \cite{Kossakowski1972,Lindblad1976}
	\begin{equation}
		\tdif{\hat{\rho}}{t} = - \ii [\hat{H}_\mathrm{H},\hat{\rho}] + \mathcal{D}[\hat{\rho}]
		\label{lindblad}
	\end{equation}
	with a Hermitian Hamiltonian that in our case is given by $\hat{H}_\mathrm{H}=\hat{H}_{\mathrm{hop}} + \hat{H}_{\mathrm{TFIM}}$ and the Lindblad dissipator
	\begin{equation}
		\mathcal{D}[\hat{\rho}] = \sum_{j,s}(\hat{L}_{j,s}\bigg(\hat{\rho}\hat{L}_{j,s}^\dagger - \frac{1}{2}\{ \hat{L}^\dagger_{j,s}\hat{L}_{j,s},\hat{\rho}\}\bigg)
	\end{equation}
	defined in terms of Lindblad operators $\hat{L}_{j,s}$ (that depend on the details of the dissipative process) and the anticommutator $\{,\}$. (For readers from soft matter physics: the Lindblad equation \eqref{lindblad} plays a similar role as the Fokker-Planck equation in classical physics, with $\mathcal{D}$ encoding friction and noise.) If we have a system in which the term $\hat{L}_{j,s}\hat{\rho}\hat{L}_{j,s}^\dagger$ can be neglected (physically corresponding to a system without particle losses) and if $\hat{L}_{j,s}=\sqrt{2\epsilon}(\hat{a}_{j,s} + \ii s \hat{a}_{j+1,s})$ \cite{AdachiTK2022}, then \cref{lindblad} can be re-written as
	\begin{equation}
		\tdif{\hat{\rho}}{t} = - \ii [\hat{H},\hat{\rho}],
	\end{equation}
	where $\hat{H}$, defined in \cref{activehamil}, is an effective Hamiltonian (that is not Hermitian).
	
\begin{figure}
		\centering
		\includegraphics[scale=0.5]{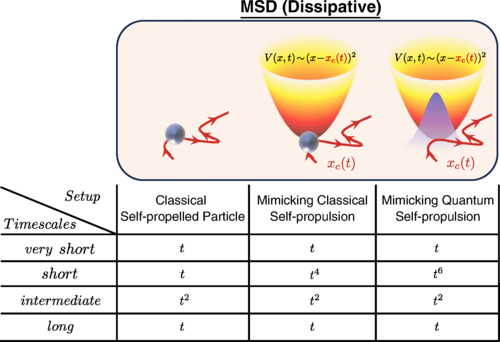}
		\caption{Mean-squared displacement of classical and quantum-mechanical active particles. (Reproduced with permission from Ref.\ \cite{ZhengL2023}.)}
		\label{fig:aqm2}
	\end{figure}

	A different way of realizing quantum active matter was proposed by \citet{ZhengL2023}. Here, activity is incorporated via a time-dependent Hamiltonian rather than via a non-Hermitian one, and the particle moves in continum space rather than on a lattice. The authors suggested that, rather than actually making a quantum-mechanical particle active, one can \textit{mimic} self-propulsion by dragging a quantum-mechanical particle along the (stochastic) trajectory $x_\mathrm{c}(t)$ of a classical active particle. Making $x_\mathrm{c}(t)$ the minimum of a harmonic trapping potential gives the Hamiltonian
	\begin{equation}
		\hat{H} = \frac{\hat{p}^2}{2m} + \frac{1}{2}m\omega (\hat{x}-x_\mathrm{c}(t))^2
		\label{quantumhamil}
	\end{equation}
	with momentum operator $\hat{p}$, position operator $\hat{x}$, mass $m$, time $t$, and frequency $\omega$. Specifically, $x_\mathrm{c}$ corresponds to the (classical but stochastic) trajectory of an active Ornstein-Uhlenbeck particle (AOUP) \cite{Szamel2014,MartinEtAl2021} and has the velocity autocorrelation
	\begin{equation}
		\braket{\dot{x}_\mathrm{c}(t)\dot{x}_\mathrm{c}(t')}=\frac{D}{\tau}e^{-\frac{|t-t'|}{\tau_\mathrm{p}}}
	\end{equation}
	with the diffusion coefficient $D$ and the persistence time $\tau_\mathrm{p}$. Dissipation is included by coupling \cref{quantumhamil} to a photon field and describing the system with the Lindblad equation (there are various possibilities for the exact structure of the dissipative term \cite{AntonovEtAl2025}). One can then compare the mean-squared displacement for a classical AOUP (\cref{fig:aqm2}, left), a classical particles dragged along the trajectories of an AOUP (\cref{fig:aqm2}, middle), and a quantum particles dragged along the trajectories of an AOUP (\cref{fig:aqm2}, right). It was found in Ref.\ \cite{ZhengL2023} that, while the results agree on very short, intermediate, and long timescales, at short timescales the mean-squared displacement is proportional to $t$ for an AOUP, to $t^4$ for the classical dragged particle and to $t^6$ for the quantum-mechanical dragged particle.
	
	Does this system count as active? Considering the criteria discussed in Section \ref{whatitis}, the one which is not obviously satisfied is directionality. After all, for an individual realization the trapping potential moves (locally has a direction) even in the absence of a particle. However, the trap motion emerges from a random process and has no direction on average (i.e. after averaging over many realizations). In this regard, it is similar to an autophoretic Janus colloid, where one finds no preferred direction when taking an average over many realizations, and distinct from that of a sedimenting colloid, where such a preferred direction exists. Therefore, we have classified this system as satisfying directionality in the table in Fig.\ \ref{criteria}, but acknowledge that one may have diverging opinions here. For this system (and for related classical ones, such as the one studied in Ref.\ \cite{FrechetteBH2024}), the classification as active, which \citet{AntonovEtAl2025} suggest,\footnote{The authors argue that they \ZT{mimic self-propulsion}, but importantly also that they \ZT{introduce a framework for engineering active quantum matter}.} requires (as always) a sufficient degree of coarse-graining. If one models explicitly both the trap and the particle, the situation is more akin to that of a passive particle driven by an active bath. However, just as one can ascribe nonreciprocal interactions to quantum nanoparticles in a modeling framework that does not explicitly account for the light source they are coupled to \cite{RieserEtAl2022,ReisenbauerEtAl2024}, the quantum system from Ref.\ \cite{ZhengL2023} possesses the dynamics of an active one on this level of description.
			
	While this section has focused on quantum-mechanical active particles, the connection should be easier to make on the level of field theories (see Section \ref{activefieldtheories}), since these can take similar forms for very different systems. There are many field-theoretical models in the open quantum system literature \cite{SiebererBMD2023,SiebererBD2016} describing at continuum level dissipative systems that violate time-reversal symmetry at local level. These are waiting for the connection with the word \ZTP{active} to be made. A good example is the study in Ref.\ \cite{YoungGFM2020}, which models a driven-dissipative quantum system described by two order parameter quantum fields $\phi_1$ and $\phi_2$ with the Langevin equations
	\begin{align}
		\zeta_1 \partial_t \phi_1&= -\Fdif{H}{\phi_1} - g_{12}\phi_1\phi_2^2 + \xi_1,\label{quantummodel1}\\
		\zeta_2 \partial_t \phi_2&= -\Fdif{H}{\phi_2} - g_{21}\phi_2\phi_1^2 + \xi_2,\label{quantummodel2}
	\end{align}
	where $\zeta_1$ and $\zeta_2$ are mobilities, $H$ is a Hamiltonian, $g_{12}$ and $g_{21}$ are coupling coefficients, and $\xi_1$ and $\xi_2$ are Gaussian white noises. The authors specifically focus on the case with $g_{12}\neq g_{21}$, in which case \cref{quantummodel1,quantummodel2} cannot be written as $\zeta_i \partial_t \phi_i = - \delta H/\delta \phi_i + \xi_i$ with some $H$ and where it shows nonequilibrium behavior. Within the active matter community, \cref{quantummodel1,quantummodel2} would probably be classified as an active field theory, specifically as a non-reciprocal Allen-Cahn equation \cite{LiuHKHKS2023} (the non-conserved variant of the non-reciprocal Cahn-Hilliard equation \eqref{nrc1}--\eqref{nrc2}).
	
	\section{Conclusion}
	In this review, we asked the question what precisely the popular term \ZTP{active matter} refers to. As the field of active matter physics has developed, more and more systems have been classified as active. Nowadays, active matter physicists study not only self-propelled colloidal particles and animals and nematic shakers, which were the historical paradigm examples for active matter, but also systems with nonreciprocal interactions and certain types of quantum systems. We have discussed in detail a number of criteria typically employed to identify systems as active, such as local energy influx used to generate motion and forces, nonequilibrium steady state behaviour, and directionality of spontaneous particle motion. Moreover, we have discussed the idea of an active field theory and the difficulties in defining what exactly this is. These issues were discussed using a number of examples from current research. The main message is that the idea of activity aims (without complete success) to single out a specific class of nonequilibrium systems, and that this classification emerges only from a coarse-grained perspective. Activity is not a strictly observer-independent or resolution-independent property of a system; its presence or absence can be a matter of viewpoint. What perhaps matters most is whether the active matter viewpoint is a helpful one for the particular system under study. One way it can be helpful is to expose similarities with other systems that might show similar emergent behaviour despite a very different microscopic prescription. This is important to keep in mind when extending the idea of active matter to novel areas, such as nonreciprocal interactions and quantum systems.
	
	We hope that our discussion also provides an introduction to what is special about active matter and what motivates studying it as a separate class of materials. Moreover, our case studies of systems only recently identified as active (dissipative systems with nonreciprocal interactions and quantum active matter) do, in passing, also provide an overview over recent developments in the field of active matter. Finally, our discussion of the term \ZTP{active matter} perhaps shows how a  search for terminological clarity, even if not completely successful, can help delineate the boundaries of an interdisciplinary research topic and clarify its links to other domains.  We expect that many of these boundaries will offer fruitful territory for future research.
	
	\acknowledgments{We thank Shikha Dhiman, Karin Everschor-Sitte, Noah Grodzinski, Zhi-Feng Huang, Robin Kopp, Sangyun Lee, Edward A. Lemke, Sarah A. M. Loos, Hartmut L\"owen, Jamir Marino, Giovanna Morigi, Alexander Morozov, Sriram Ramaswamy, Janik Sch\"uttler, Jan Smrek, Anton Souslov, Thomas Speck, Jann van der Meer, Artur Widera, Sam Wilken, Sina Wittmann, and Raphael Wittkowski for helpful discussions. M.t.V.\ is funded by the Deutsche Forschungsgemeinschaft (DFG, German Research Foundation) -- SFB 1551, Project-ID  464588647. This study contributes to research done in the Mainz Institute of Multiscale Modeling, M$^3$ODEL.}

\end{document}